\documentclass[11pt]{article}
\usepackage{epsfig, amsfonts, amsmath, amssymb, fullpage}

\usepackage{epstopdf}

\usepackage{algorithmic}
\usepackage[nothing]{algorithm}

\usepackage{float}
\usepackage{subfig}
\usepackage{multirow}

\newcommand{\real}{\mathbb{R}}

\newcommand{\len}{\textsf{spath}}
\newcommand{\Prob}{\mathbb{P}}

\title{\Large Multiscale Network Generation}
\author{
Alexander Gutfraind\thanks{Center for Computational Biology and Bioinformatics, University of Texas at Austin, 1 University Station, C0930, Austin, TX 78712 \href{mailto:agutfraind.research@gmail.com}{\tt agutfraind.research@gmail.com}}
\and
Lauren Ancel Meyers\thanks{Section of Integrative Biology, University of Texas at Austin, 1 University Station, C0930, Austin, Texas 78712; Santa Fe Institute, Santa Fe, NM 87501
\href{mailto:laurenmeyers@austin.utexas.edu}{\tt laurenmeyers@austin.utexas.edu}}
\and
Ilya Safro\thanks{Mathematics and Computer Science Division, Argonne National Laboratory, 9700 S Cass Ave, Argonne, IL 60439 \href{mailto:safro@mcs.anl.gov}{\tt safro@mcs.anl.gov}}
}

\date{\normalsize \today}
\begin{document}

\maketitle

\begin{abstract}
Networks are widely used in science and technology to represent relationships between entities, such as social or ecological links between organisms, enzymatic interactions in metabolic systems, or computer infrastructure. Statistical analyses of networks can provide critical insights into the structure, function, dynamics, and evolution of those systems.
However, the structures of real-world networks are often not known completely, and they may exhibit considerable variation so that no single network is sufficiently representative of a system.
In such situations, researchers may turn to proxy data from related systems, sophisticated methods for network inference, or synthetic networks. 
Here, we introduce a flexible method for synthesizing realistic ensembles of networks starting from a known network, through a series of 
mappings that coarsen and later refine the network structure by randomized editing.
The method, MUSKETEER,
preserves structural properties with minimal bias, including unknown or unspecified features, while introducing realistic variability at multiple scales.  
Using examples from several domains, we show that MUSKETEER produces the intended stochasticity 
while achieving greater fidelity across a suite of network properties than do other commonly used network generation algorithms.

\end{abstract}

\section{Introduction}

Across the sciences, medicine, and engineering, networks are widely used to represent connections between entities, because they 
provide intuitive windows into the function, dynamics, and evolution of natural and man-made systems \cite{Newman:2010:NI:1809753}.
However, high-quality, large-scale network data is often not available, because of economic, legal, technological, or other obstacles \cite{chakrabarti2006graph,brase2009modeling}. 
For example, the human contact networks along with infectious diseases spread are notoriously difficult to estimate, 
and thus our understanding of the dynamics and control of epidemics stems from models that make highly simplifying assumptions or simulate contact networks from incomplete or proxy data  
\cite{Eubank04,Keeling_Rohani_2008,Meyers05}. In another domain, the development of cybersecurity systems requires testing across diverse threat scenarios and validation across diverse network structures that are not yet known, in 
anticipation of the computer networks of the future \cite{dunlavy2009mathematical}. In both examples, the systems of interest cannot be represented by a single exemplar network, 
but must instead be modeled as collections of networks in which the variation among them may be just as important as their common features.
Such cases point to the importance of data-driven methods for synthesizing networks that capture both the essential features of a system and realistic variability in order to use them in such tasks as simulations, analysis, and decision making. 

A good synthetic network must meet two criteria. First, it should be realistic with respect to structural features that govern the domain-specific processes of interest (system function, dynamics, evolution, etc.). Consider the following examples: 
\begin{itemize}
\item Models of social networks should be able not only to reproduce structural features such as small-world properties, but also, and perhaps more important, to emulate emergent sociological phenomena such as  interactions between individuals in a community, as driven by their psychological needs and daily routines.  That is, the generated network should show similar interactions by its artificial individuals, as determined by implicit psychological and social rules.
\item Models of connected solar energy collectors of different sizes and capacities should simulate realistic energy outputs influenced by the weather.
\item Models of metabolic interactions should ultimately reflect biochemical properties of a cell. 
\end{itemize}
Second, a synthetic network should reflect naturally occurring stochasticity in a system, without systematic bias that departs from reality. This feature is particularly important for benchmarking and evaluating the robustness of network-based algorithms, anonymizing networks, and generating plausible hypothetical scenarios.

A number of network generation methods have been developed, and these fall into two classes: generative models and editing methods (see surveys in \cite{chakrabarti2006graph,brase2009modeling,dunlavy2009mathematical}).
The first set of methods produce (by using randomization and replication) a graph from a small initial seed graph (sometimes empty). The goal of such generation is to produce the  structure that matches real data in \emph{prespecified} properties, such as
the degree distribution \cite{Albert2002,Newman03thestructure,mahadevan2006systematic}, clustering \cite{Bansal09}, and the number of small subgraphs \cite{robins2007introduction}.
These methods are attractive because they often produce networks with the desired features and are grounded in well-developed theory (e.g., \cite{erdos1960erg}).
Some of these graph generators mechanistically model network growth \cite{barabasi,krapivsky,Leskovec08}, whereas others incorporate domain-specific information such as geographic location \cite{Watts1998} and cyber networks topological properties \cite{medina,dunlavy2009mathematical}. 
One of the most successful generative strategies is based on Kronecker graphs \cite{kronmodel} (including stochastic Kronecker graphs; see also related work \cite{chakrabarti2006graph,palla2010multifractal}). Graphs generated by this model preserve properties such as degree distribution, diameter, eigenvalues, and eigenvectors. Such methods often describe an evolutionary process that can potentially lead to the original network; however, the probability that it will lead to the structure that is approximately isomorphic to the original one is usually negligible. This makes generative methods ill-suited for studies such as simulations when one may need to work with systems that are similar to the original.
The other class of network generators, graph editing \cite{mihail2003markov}, 
starts with a given (real or empirical) network and randomly changes its components until the network becomes sufficiently different from the original network. 
These are designed to introduce variability while preserving key structural properties. 

Because networks are used in many diverse scientific and technological domains, constructing a network generator suitable for all of them can present significant challenges.
Indeed, although various models and algorithms have been proposed for synthesizing networks, 
our analysis suggests that they synthesize networks that fall short of what is desirable (see Appendix \ref{sec:app-alternatives}).
Existing methods (1) involve generative mechanisms or system-specific assumptions that are not plausible across domains, (2) focus on one or a few predefined topological features, such as degree distribution and clustering, at the expense of others that may be critical but are perhaps unrecognized or cannot be incorporated into the generator easily, and/or (3) reliably reproduce a set of target properties but fail to capture naturally occurring stochasticity in those properties. 
Here, we introduce an editing-based approach that makes progress in meeting some of these challenges and opens a new research direction in this class of methods. Starting from a single known or hypothesized network from any domain, it synthesizes ensembles of networks that preserve, on average, a diverse set of structural features at multiple scales, including degree, other measures of centrality, degree assortativity, path lengths, clustering, and modularity, while introducing unbiased variability across the ensemble in many of these properties. 

The core methods are inspired by applications of the theory of multiscale methods   \cite{brandt:review01,briggs,mgbooktrott} to combinatorial optimization \cite{brandt:optstrat,safro:relaxml}. 
The original method starts with a complex optimization problem represented on a network, 
constructs a hierarchy of decreasing in size networks $G_0=G,G_1,...,G_k$ via a coarsening procedure \cite{safro:relaxml}, 
solves a simpler optimization problem at the coarsest scale, and then iteratively uncoarsens the solution by gradual projections from coarse to next-finer scales.
Similarly, we create a hierarchy of coarse networks; but, in contrast to the multiscale methods for computational and optimization problems, we do not optimize anything but edit the network at all scales of coarseness. 
During the editing process we allow only local changes in the network that will be the results of local decisions only. In other words, the problem of network editing is formulated and solved at all scales where primitives at the coarse scale (such as coarse nodes and edges) represent aggregates of primitives at previous finer scale. Analogous to multiscale methods for computational problems  \cite{brandt:review01,brandt:optstrat}, by using appropriate coarsening we are able to detect and use the ``geometry'' behind the original network at multiple scales, which can be interpreted as an additional property that is not captured by other network generation methods. 
Moreover, it is known that the topology of many complex networks is hierarchical and thus might be produced through iterations of generative laws at multiple scales. In general, such generative laws often can be different at different scales, as evidenced by the finding that complex networks are self-dissimilar across scales \cite{Carlson02, Itzkovitz05, palla2005uncovering, Wolpert07, Binder08, Mones12}. These differences can naturally be reflected in the proposed multiscale framework.
The multiscale framework also helps us achieve linear running time in terms of the number of edges in the original network. 

This method has been implemented in the software tool Multiscale Entropic Network Generator (MUSKETEER), which is freely available \cite{musketeer:page}. In the following sections we describe the procedure and evaluate its performance in comparison with other network generation methods on networked systems from several different domains. 
In Section~\ref{sec:mult_strategy} we describe the main multiscale strategy and our algorithm.
In Section ~\ref{sec:exper} we present an experimental evaluation of the proposed strategy on several types of networks. In Section \ref{sec:discussion} we discuss some of the advantages and possible limitations of MUSKETEER. In Section \ref{sec:concl} we briefly present our conclusions.
The appendices discuss the technical details of the algorithm, show additional performance data, demonstrate the strengths and weaknesses of existing approaches, and present illustrations of generated networks.

\section{Multiscale Network Generation}\label{sec:mult_strategy}
\begin{figure}[htb]
\begin{center}
\includegraphics[width=0.6\textwidth]{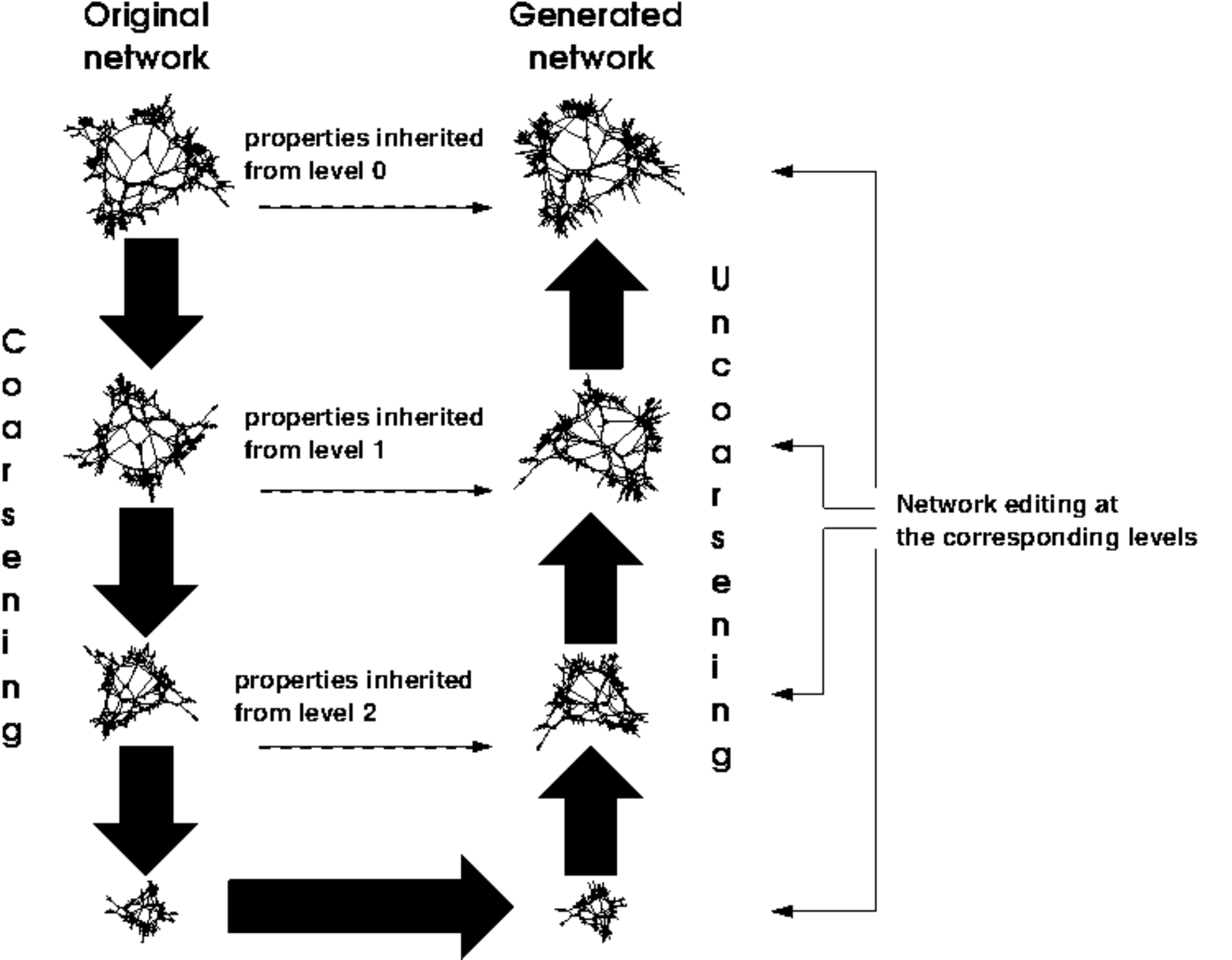}
\end{center}
  \caption{The V-cycle scheme underlying multiscale network generation.
The original network is first coarsened into smaller and smaller networks, and then the process is reversed.
At each scale the network receives a controlled perturbation, ultimately generating a new network patterned on the original. These perturbations take into account properties of the coarse networks from the corresponding scales and preserve similarities with them if needed (dashed arrows).   
The number of levels is usually greater than the four shown here and depends primarily on the structure and size of the original graph. 
\label{fig:illustration}}
\end{figure}

The network generation process is inspired by the algebraic multigrid V-cycle scheme (a multiscale method for solving  linear algebra problems by using coarsening-uncoarsening iterations) where the network's Laplacian is taken through multiple coarsening steps, followed by multiple uncoarsenings that take it back to the original scale \cite{brandt:optstrat,safro:relaxml}. 
(A pseudocode of the multiscale algorithm is presented in Algorithm \ref{alg:M}, and its flowchart is in Figure \ref{fig:illustration}.)
To generate a new network, we start from a \emph{known} network $G_0 = G$, which may be empirical (estimated directly from data) or otherwise grounded in knowledge of the system. 
The network is coarsened repeatedly in order to create the sequence of graphs $\{G_i\}_{i=1}^k$ ($V_i$ and $E_i$ will correspond to the sets of nodes and edges of $G_i$, respectively).
Each coarsening involves computing a projection of the graph Laplacian:
\begin{equation}\label{eq:amg1}
L_{i+1} = (P_i)^T L_i P_i,
\end{equation}
where $L_i$ is the Laplacian matrix of $G_i$ and
the $P_i \in \real^{|V_i|\times |V_{i+1}|}$ is the restriction matrix that describes the strength of connectivity between $i$th-level (fine) nodes 
and corresponding aggregates that form coarse nodes (Figure \ref{fig:illustration}, additional details in Appendix \ref{sec:app-technical}).
Coarsening terminates when changes at coarser levels are not requested by the user or if the coarsened network is already very small or very dense. 
At the coarsest network $G_k$, we apply an editing procedure to obtain $\tilde{G}_{k}$ (Algorithm ~\ref{alg:M}, line 2).
Given the edited network $\tilde{G}_{i+1}$ at level $i+1$ we uncoarsen it to obtain the level $i$ network $G'_{i}$.
Uncoarsening of unedited aggregates is a straightforward reverse projection operation;
edited and new aggregates are interpolated by resampling (i.e. copying) nodes and edges from ${G}_{i}$. 
Then $G'_{i}$ is edited into $\tilde{G}_{i}$ which completes the revision process at level $i$ (Algorithm \ref{alg:M}, line 7).
The edit-and-uncoarsen process is repeated all the way up until the original level is reached. 
At that point the final result is generated: a new network $\tilde{G}_{0}$.

\begin{algorithm}[tbh]\label{alg:main}
    \caption{$\textsf{ReviseGraph}(G_i)$ - the main block of MUSKETEER}
\begin{algorithmic}[1]\label{alg:M}

\IF {$G_i$ is too dense, too small or there are no edits at coarser levels}
    \STATE \textbf{Return} $\textsf{EditEdgesAndNodes}(G_i)$
\ENDIF
\STATE $G_{i+1} ~ \leftarrow ~ \textsf{Coarsen}(G_i)$
\STATE $\tilde{G}_{i+1} ~ \leftarrow ~ \textsf{ReviseGraph}(G_{i+1})$
\STATE $G'_i ~ \leftarrow ~ \textsf{Interpolate}(\tilde{G}_{i+1})$
\STATE $\tilde{G}_i ~\leftarrow ~ \textsf{EditEdgesAndNodes}(G'_i)$
\STATE $\tilde{G}_i ~\leftarrow ~ \textsf{UserDefinedAdjustment}(\tilde{G}_i)$
\STATE \textbf{Return} $\tilde{G}_i$
\end{algorithmic}
\end{algorithm}

The function $\textsf{EditEdgesAndNodes}$ applies intentionally primitive changes of inserting and deleting single nodes and edges, but the location of those edits is carefully controlled.
Before making the edits, $\textsf{EditEdgesAndNodes}$ estimates the distribution of a quantity termed the second shortest path length, $\len$:
Given an edge $\{u,v\}$, $\len \{u,v\}$ is the number of edges in the shortest $u\to v$ path that does not include $\{u,v\}$, if such a path exists.
The distribution of $\len$ is estimated by sampling random edges $R\subset E_i$ in the graph $G_i$, namely,
\begin{equation}
\Prob_{G_i}[d] \approx \frac{|\{ \{u,v\}\in R ~:~ \len \{u,v\} = d\}|}{|R|}.
\end{equation}
Given $\Prob_{G_i}[d]$, new edges are inserted by sampling $d$ from it, choosing a random node $u$, and inserting an edge to a node $v$ at distance $d$ from $u$.
This sampling process tends to maintain multiple structural properties of the graph.
For example, the clustering coefficient is preserved because in clustered graphs the probability of $d=2$ is elevated and the edge $\{u,v\}$ above would construct a triangle.
Other distance-based properties such as the average geodesic distance are also preserved under this sampling
because the probability of inserting an edge with a large $d$ is low 
(if inserted, such an edge would significantly reduce the average distance. See Section \ref{sec:app-extended-perf} for performance results).
When new nodes are inserted, their degree is randomly chosen based on the degree of an existing node in the network,
in a kind of network \emph{resampling}.
The new node is connected to a random existing node, and then subsequent edges from it are inserted by sampling from $\Prob_{G_i}[d]$.
Deletion of nodes and edges occurs through random sampling (but see Section \ref{sec:app-technical} for optional features). 
The number of such edits (i.e., the edit rate) can be controlled by the user and can also be tuned for a particular class of networks.

\begin{figure}[H]\begin{center}
\subfloat[Fine-level edge edit]{\label{fig:gr2-shallow-edge}\includegraphics[width=0.5\textwidth]{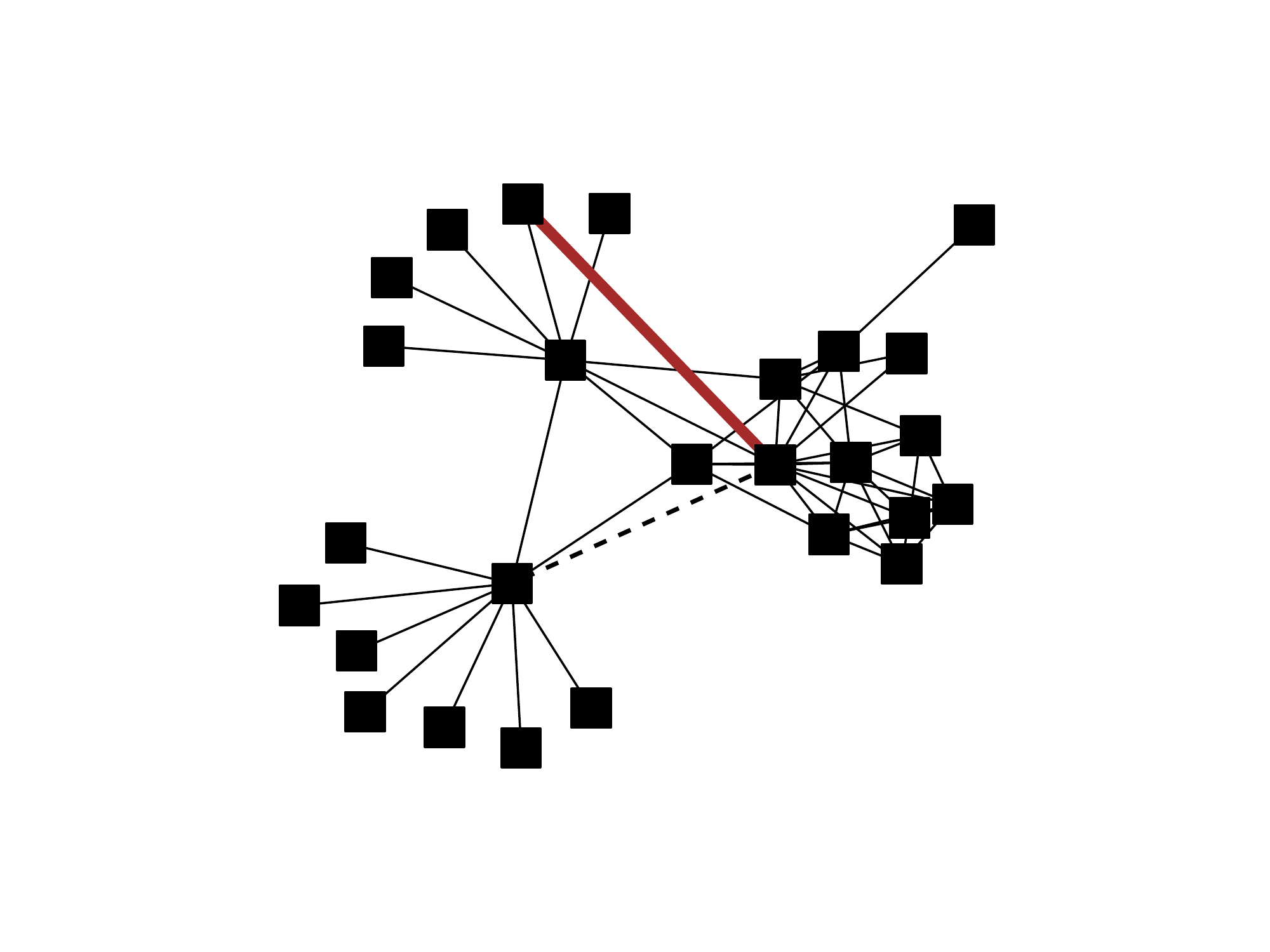}}
\subfloat[Coarse-level edge edit]{\label{fig:gr2-deep-edge}\includegraphics[width=0.5\textwidth]{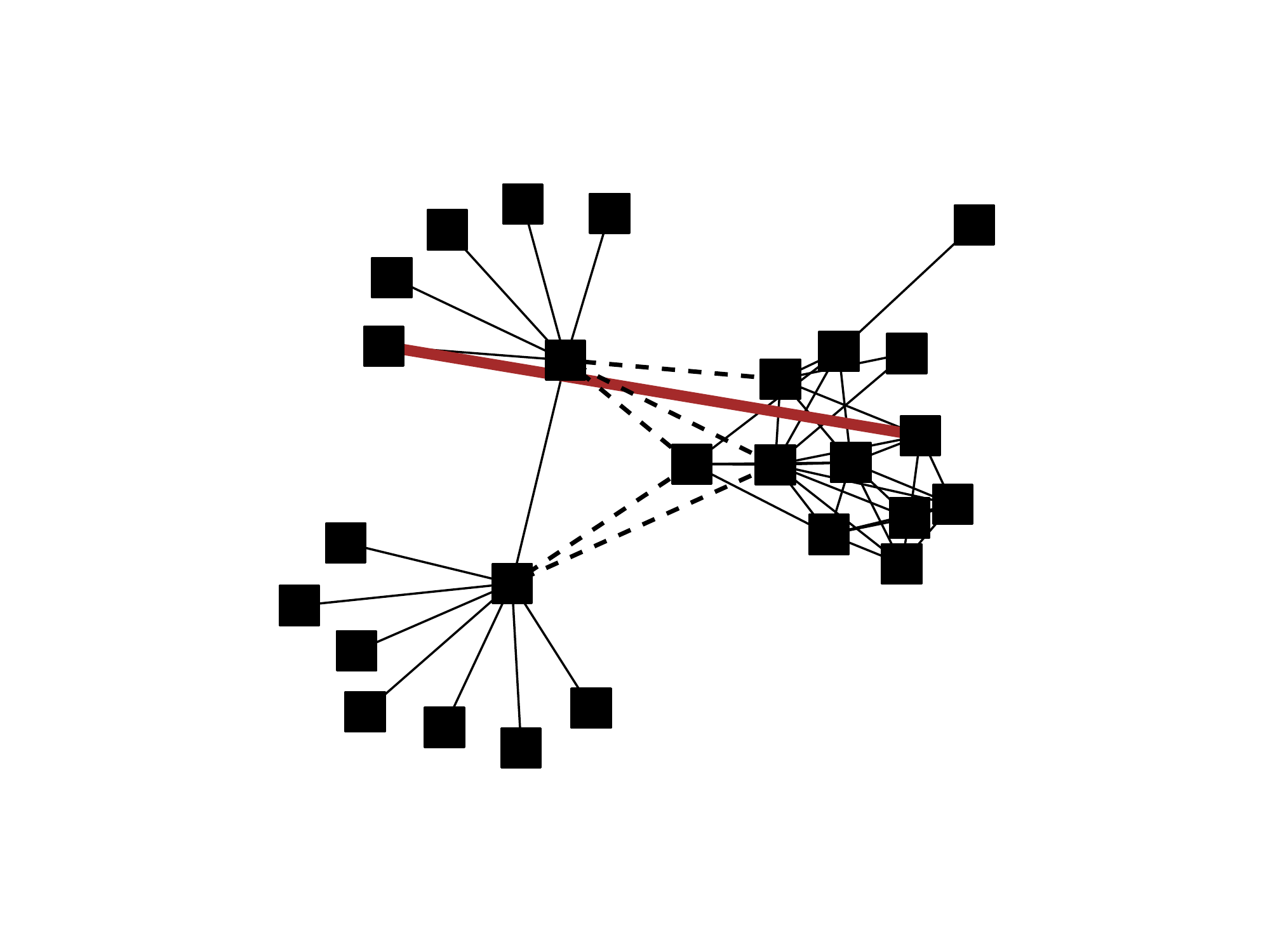} } \\
\subfloat[Fine-level node edit]{\label{fig:gr2-shallow-node}\includegraphics[width=0.5\textwidth]{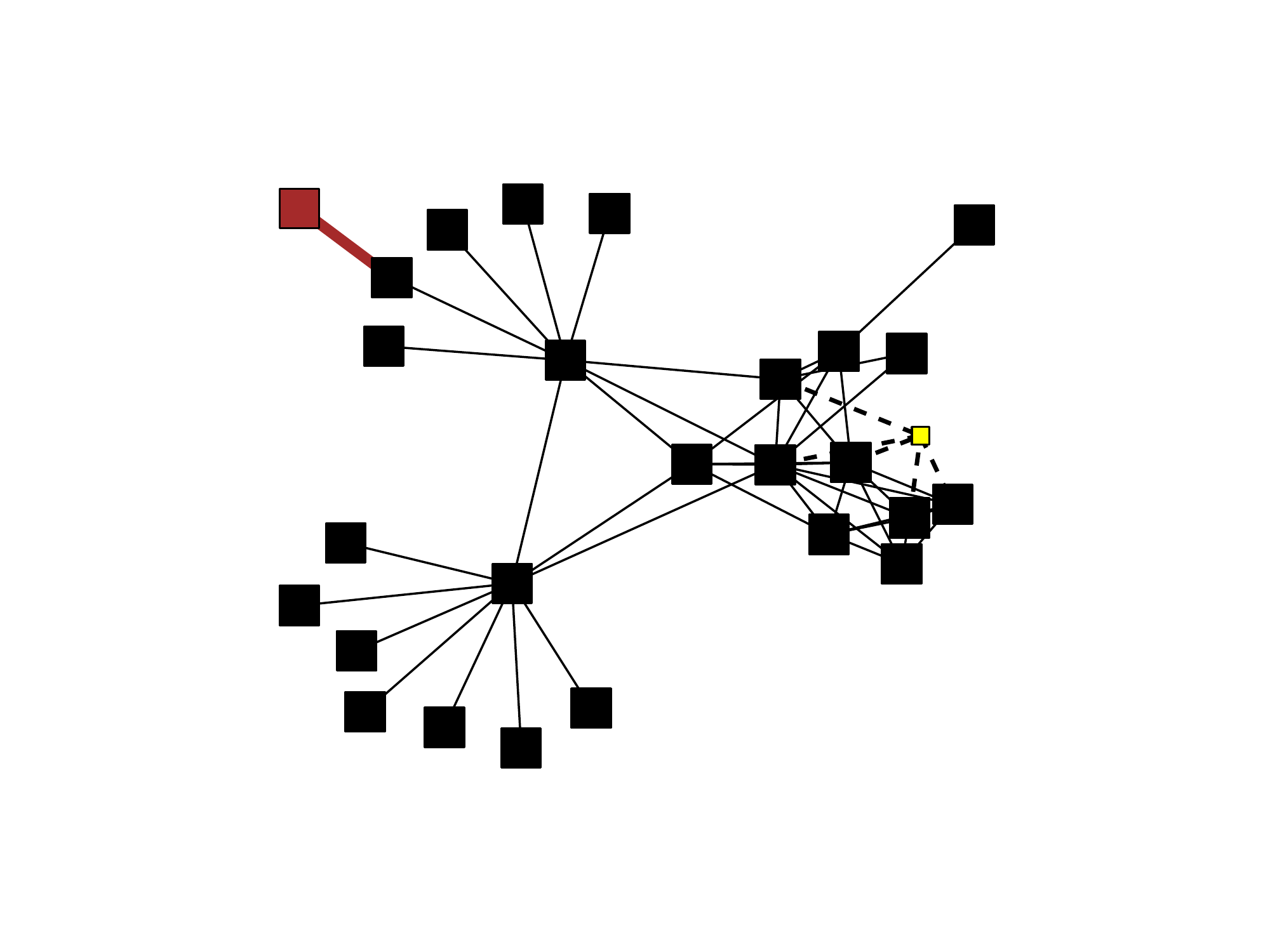}}
\subfloat[Coarse-level node edit]{\label{fig:gr2-deep-node}\includegraphics[width=0.5\textwidth]{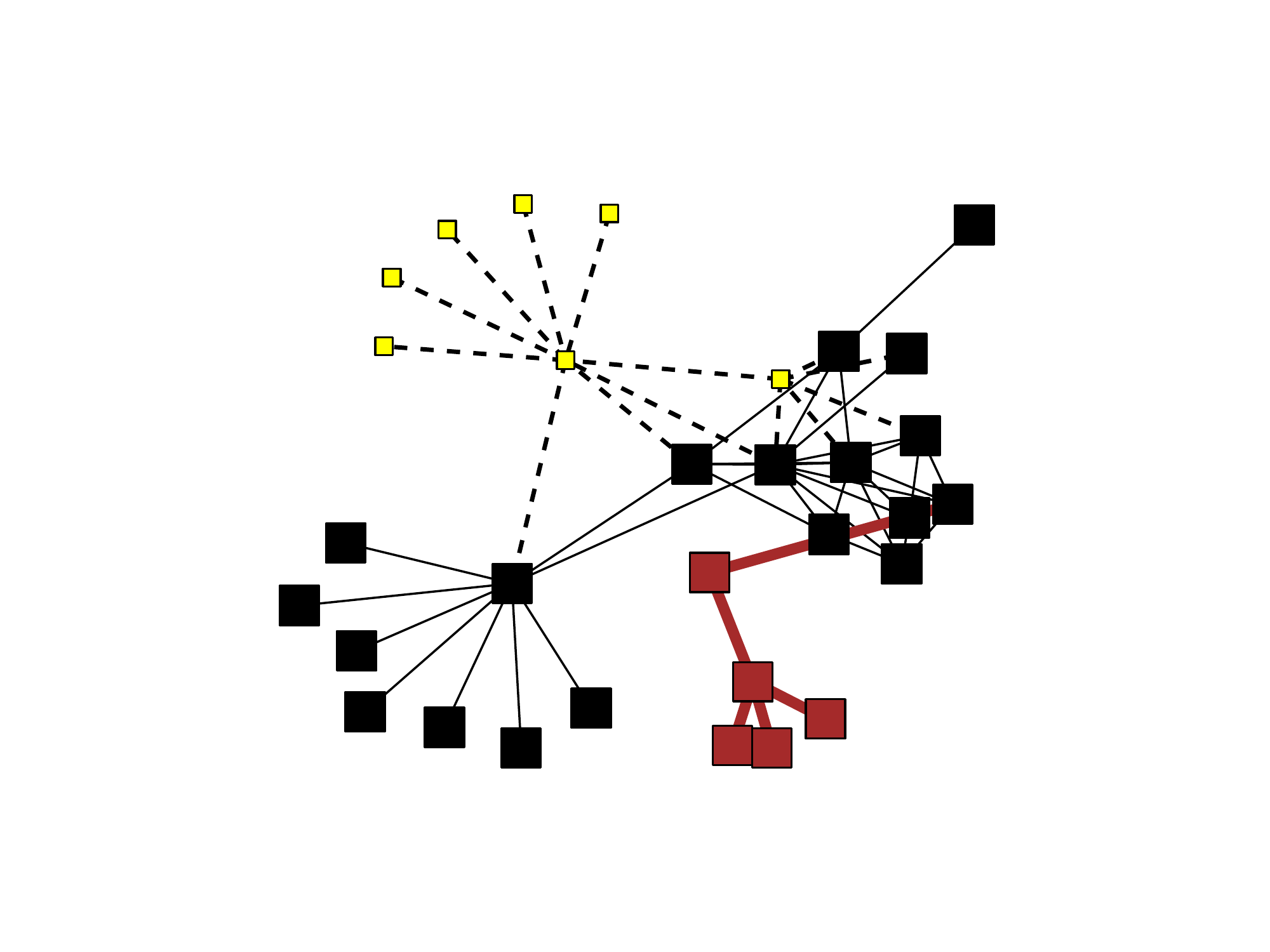}}
\end{center}
\caption{Effect of editing depth on the resulting replica.
\label{fig:edits} (a) Fine-level edit to edges causes rewiring of individual edges.
(b) Coarse-level edit to edges breaks or adds connections between groups of nodes.
(c) Fine-level edit to nodes creates and deletes single nodes and their connections.
(d) Coarse-level edit to nodes creates new clusters of nodes together with their connections.
The original graph is modeled on a communication network between autonomous systems on the Internet.
Original nodes and edges are in black; new nodes and edges are in red; deleted nodes are in yellow; deleted edges
are in dashed lines.
}
\end{figure}

The editing process performs only simple operations, but those operations can generate high-entropy changes because they could be applied everywhere in the entire hierarchy $\{G_i\}_{i=0}^k$.
As a result, when nodes are added to the finest levels of coarsening, they have the effect of just adding nodes to the replica;
but when this same addition occurs at coarser (i.e., deeper) levels, the new nodes will correspond to several nodes or large communities in the final replica.
The deeper the graph in which the change occurs, the more significant is change to the topology.
In this way, one can make large, nontrivial changes to the topology of the replica through a simple random process at deeper levels.
Usually the creation of a node is accompanied by the creation of edges to it, and the ultimate effect of those edges on the replica
depends on the level of coarsening:
at the finest level, edges are just edges, whereas at deeper levels of coarsening an edge lead to the creation
of many interconnections across entire communities of aggregated fine nodes (see Figure~\ref{fig:edits}).
MUSKETEER applies the same editing process at all scales, unless the user adds new adjustment requirements (line 8 in Algorithm \ref{alg:M}) such as control of some invariant (like connectivity of the graph).

Additional details about the algorithm are given in Appendix~\ref{sec:app-technical}.

MUSKETEER is expected to produce realistic replicas because it reproduces the original
network everywhere except at the edited components, and the edited components are simply resampled from the original network.
As the edit rate is lowered, the replicas become arbitrarily close to the original in every respect, including properties that are unknown or unspecified, 
provided the property is not highly sensitive to small perturbations of the network. Furthermore, since all edits in MUSKETEER are 
constrained by the properties of the network at the appropriate structural level, 
the replicas are expected to preserve scale-dependent features and can be self-dissimilar across scales (in accordance with the original network).
In contrast, random graph model generators are often unable to reproduce unspecified and scale-dependent properties, as demonstrated below.

\section{Performance of the Generator}\label{sec:exper}

We evaluate MUSKETEER in terms of both the fidelity and the variability of the replicas, using an empirical example network from infectious disease epidemiology and two 
well-studied random graph networks (see Appendix \ref{sec:app-extended-perf} for examples from other domains, including proteomics, energy, and linguistics). A number of network properties are known to influence the dynamics of infectious disease outbreaks, including 
the number of nodes (individuals); the degree distribution (contacts) \cite{Newman03thestructure,Meyers05}; 
clustering coefficient (number of triangles out of all possible triangles) \cite{Bansal09,volz2011effects};
degree assortativity (tendency for nodes to connect to others with similar degree) \cite{Newman03thestructure,Newman03mixing}, which 
we measure according to~\cite{doyle2005robust,li2005towards}; 
node centrality~\cite{Newman03thestructure}; average distance, and modularity as measured by~\cite{Blondel08}, which can give rise
to multi-wave epidemics~\cite{Galstyan07modular}.
Many of these properties are also highly relevant to system function and dynamics in other domains~\cite{Gutfraind10}.

\begin{figure}[htb] \begin{center}
\subfloat[]{\label{fig:potterat-fig}\includegraphics[width=0.5\textwidth]{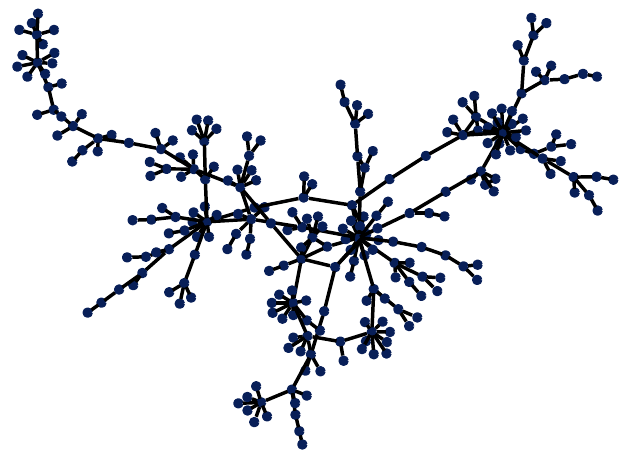}}
\subfloat[]{\label{fig:potterat-rep-fig}\includegraphics[width=0.5\textwidth]{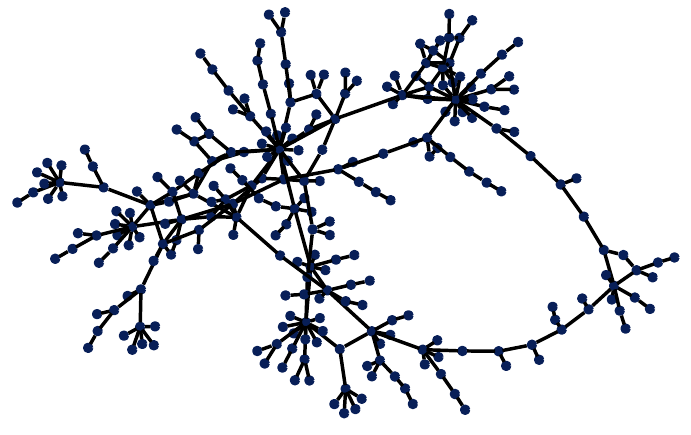}}
\end{center} 
\caption{Replica of an empirical sexual contact network. (a) The original contact network estimated from \cite{Potterat02} and (b) its replica generated by MUSKETEER.
The replica has over $30\%$ synthetic nodes and edges, yet still visually resembles the original network.
\label{fig:musk-illustration}}
\end{figure}

One particularly well-studied epidemiological network is based on data collected in Colorado Springs by Potterat et al. (\cite{Potterat02} Fig.~2A).
It contains 250 individuals who were in contact in the 1980s through sex or injection drug use.
In our experiments we edited $8\%$ of the nodes at the finest level and $7\%$ on the next coarsest level. 
The same edit rates were assigned to the edges at those levels.
Thus, crudely, the total edit rate is $30\%$; but it is actually higher because the topological effect of coarse-level edits is larger than their rate might suggest, and is approximately doubled with each level. 
Even at this conservative estimate of editing, over $30\%$ of the nodes in the replicas are synthetic, and over $50\%$ of the edges are (changes to the nodes immediately change all incident edges).
This extreme randomization is chosen only to demonstrate the ability of the method to generate high-realism networks despite the high editing.
The network and a replica of it are shown in Figure~\ref{fig:musk-illustration}.
In practical applications the appropriate edit rate should be much lower and determined by the application:
the edit rate should be high only when the input network is highly different from other networks in its class.

\begin{figure}[htb] \begin{center}
\subfloat[]{\label{fig:potterat}\includegraphics[width=0.5\textwidth]{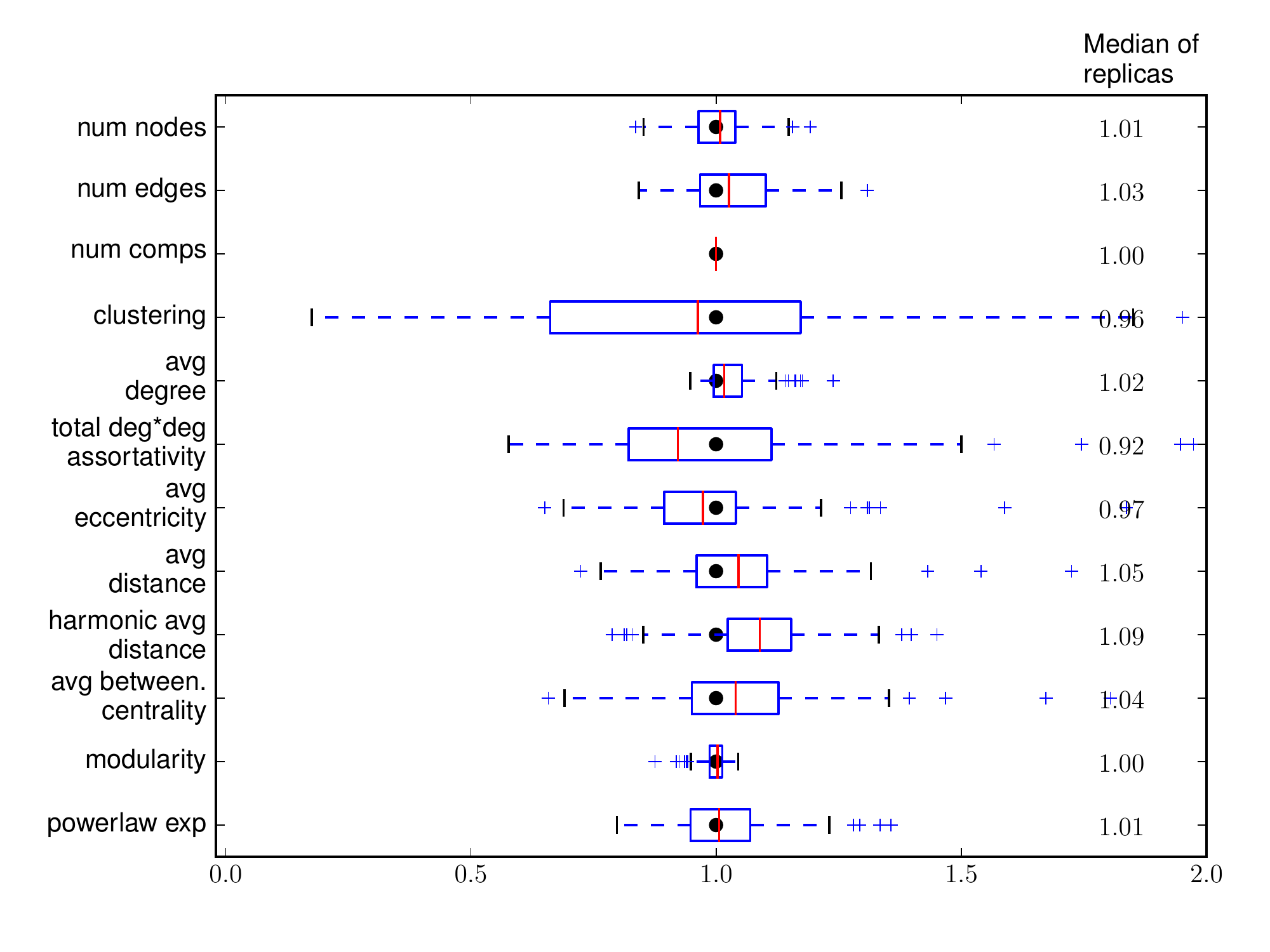}}%
\subfloat[]{\label{fig:potterat-deg}\includegraphics[width=0.5\textwidth]{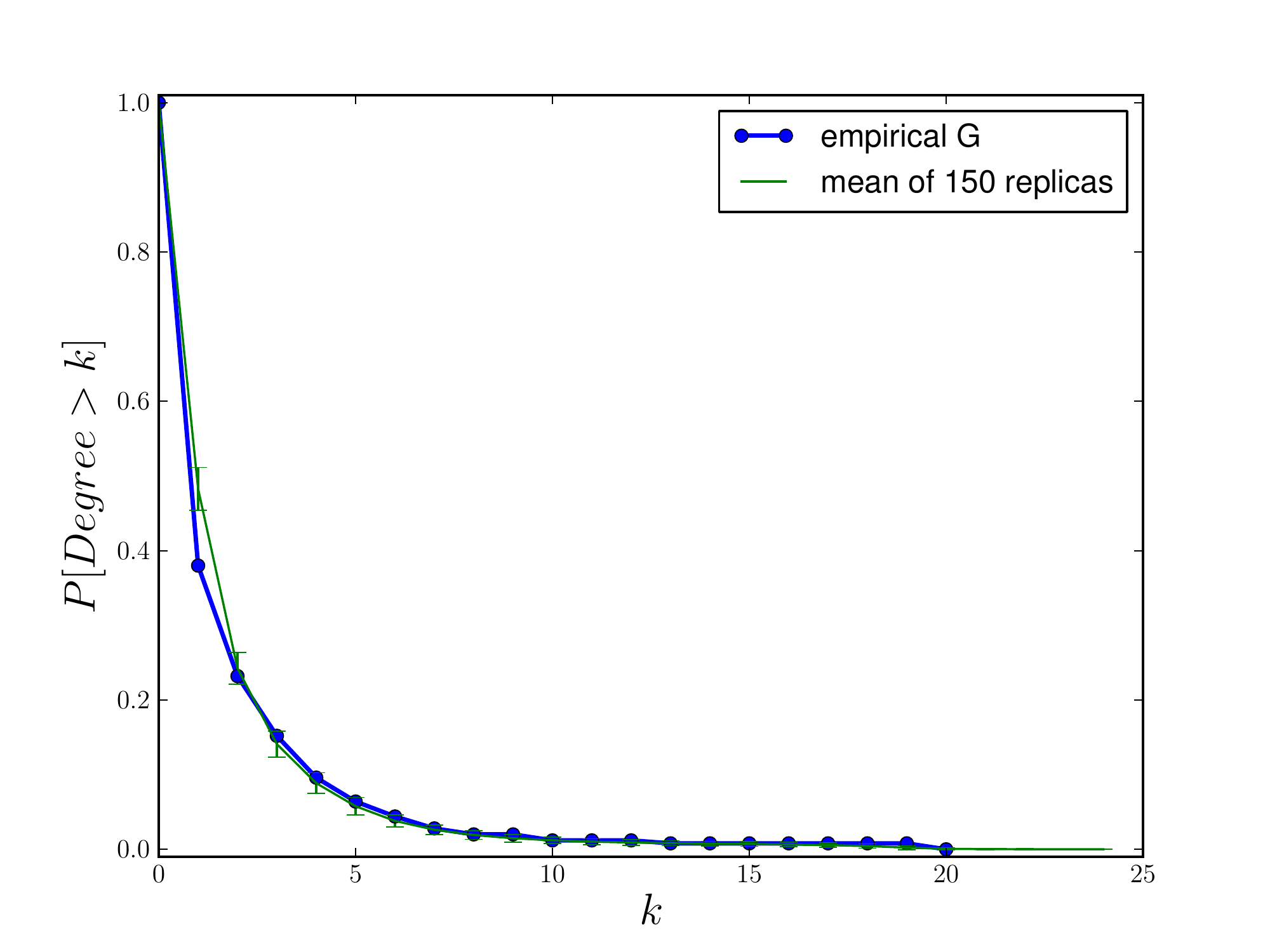}}
\end{center}
\caption{Performance of MUSKETEER on a sexual contact network (\cite{Potterat02}). 
(a) Structural properties of 150 generated networks (boxplots) compared with the original network (black dots).
All properties of the replicas are normalized so that the value in the original network is mapped to 1.0.
The standard boxplots indicate the middle 50 percent of the data (from the 25th to 75th percentile), and whiskers extend to points in either direction
whose values are 50\% greater than the range of values inside the box (interquartile range).
Generally, the replicated networks reflect the properties of the original network with unbiased variation around the median. 
(b) Average cumulative degree distribution of generated and original networks. Graph shows average degree distribution across replicas 
(green line and error bars) 
compared to original network (blue lines and points). Error bars indicate one standard deviation for replicas.
\label{fig:firststats}}
\end{figure}

To evaluate the performance of MUSKETEER, we generated a large number of replicas and compared them to the original network for a variety of local and global structural properties (Figure~\ref{fig:firststats}a). 
For most properties, the ensemble yields a median value close to the original value and range of values that is fairly symmetric about the median. The degree distribution as a whole is preserved 
with slight variation across replicas (Figure~\ref{fig:firststats}b)

\begin{figure}[htb] \begin{center}
\subfloat[Replication of an  Erd\H os-R\'enyi Graph]{\label{fig:er}\includegraphics[width=0.5\textwidth]{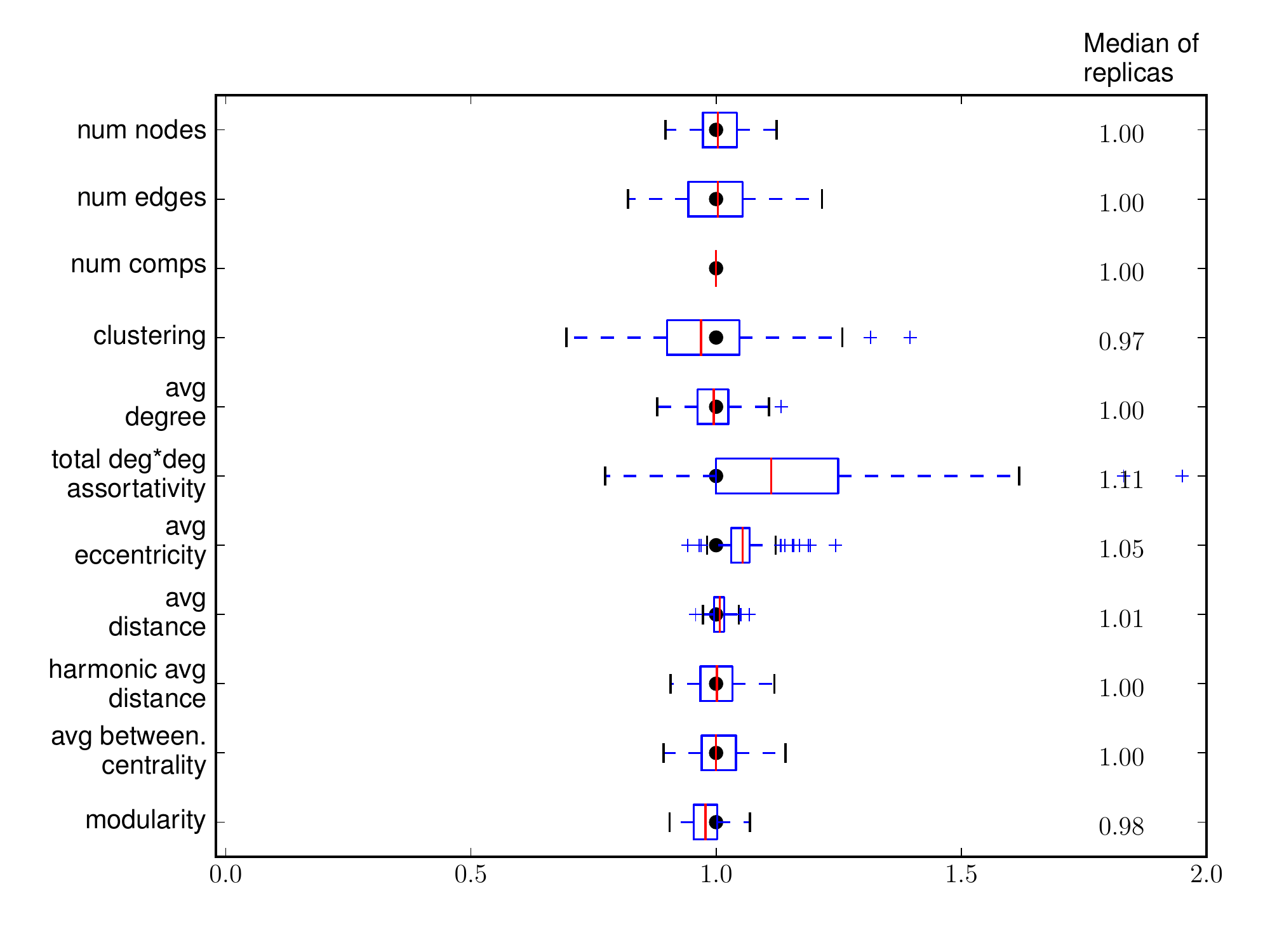}}
\subfloat[Replication of a Bar\'abasi Albert Scale-Free Graph]{\label{fig:ba}\includegraphics[width=0.5\textwidth]{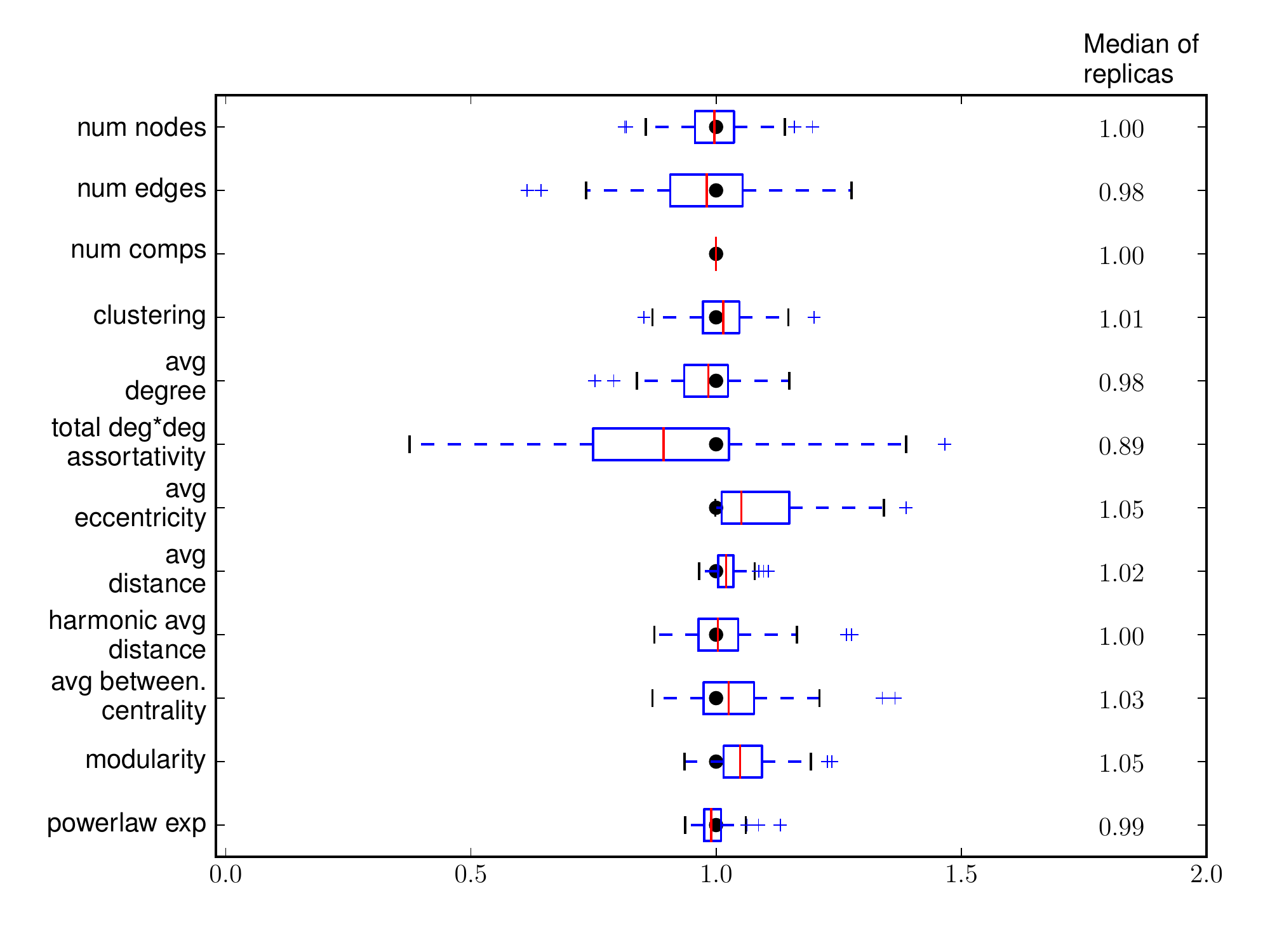}}
\end{center}
\caption{Comparison of two network models with 150 replicas.
Lines show the properties of individual networks (black dots: original, boxplot: replicas).
In most properties the replicated networks are distributed about the original
with desirable variation.
There is little or no bias in the properties of the replicas despite the fact that
at least $50\%$ of the edges in those networks are new.
The generation parameters are $p=0.05$ for ER and $m=10$ in BA. 
\label{fig:parallel-2models}}
\end{figure}

We also applied MUSKETEER to synthesize ensembles of networks starting from 300-node networks generated under
two well-understood random graph models: 
the Erd\H os-R\'enyi (ER) model, which yields simple random networks with binomial degree distributions (Poisson in the limit of large networks)~\cite{erdos1960erg}, 
and the Barab\'asi-Albert (BA) preferential attachment model, which yields scale free networks, with power law degree distributions~\cite{barabasi}.
As with the sexual contact network, MUSKETEER produced network ensembles with high fidelity and a reasonable degree of unbiased variability (Fig.~\ref{fig:parallel-2models}).

We used the sexual contact network to compare the performance of MUSKETEER with that of other widely used network generation models, such as the ER and BA models~\cite{erdos1960erg,barabasi}, the configuration model preserving the original degree sequence,
the Watts-Strogatz small-world model, the Kronecker random graph model, and single rounds of edge edits and edge swaps (results presented in Appendix \ref{sec:app-alternatives}).
MUSKETEER is the only method that was able to preserve all the structural features considered in Figure~\ref{fig:musk-illustration} 
while introducing unbiased variability across replicas. 
In general, the other methods achieve high fidelity only for properties explicitly targeted by the method and, in some cases, without introducing the desired variability. 
For example, while the configuration model correctly reproduces the average degree of the network, its replicas poorly match
the Colorado Springs network in other important properties, including clustering, which is too low, and the degree-degree product, which is too high.
The advanced randomized Kronecker graph approach produces networks with similarly lowered clustering.

Synthetic networks should not only reflect realistic structures at multiple scales  
but also capture the fundamental function, dynamics, or evolution of the system. That is, we
must return to the initial motivation for synthesizing ensembles of realistic networks
and ask whether they are realistic with respect to our higher-level objectives. For example, 
we might ask whether synthetic epidemiological contact network yield realistic epidemics.
We assess the performance of MUSKETEER in this respect on the sexual contact network described above. 
This network has been used by infectious disease epidemiologists to investigate the spread of sexually transmitted 
diseases and evaluating public health interventions in this population. One can imagine 
synthesizing variations on that network in order to study transmission dynamics and control 
across a range of plausible populations. Thus, we ask whether the replicas produced by MUSKETEER give rise
to similar epidemiological dynamics, in terms of the size and timing of a disease outbreak simulated on the network.
We compare the epidemiological realism of the MUSKETEER replicas with networks generated by using the configuration model~\cite{Chung02}
preserving the degree distribution and the BA scale free model ~\cite{barabasi} calibrated to preserve the density of the graph. 
Both models have been used extensively to model epidemiological processes~\cite{Meyers05,Newman:2010:NI:1809753} 
Specifically, we simulate a susceptible-exposed-infected-recovered (SEIR) model~\cite{keeling2008modeling} on the original and replica networks.
For simplicity, we use discrete time steps of $1$ day and that infected individuals have a latent period of $2$ days ($\pm 1$)
followed by an infectious period of $9$ days ($\pm 1$).
Each infectee has a $50\%$ chance per day of infecting each susceptible neighbor.

\begin{figure}[!htb] \begin{center}
\includegraphics[width=0.7\textwidth]{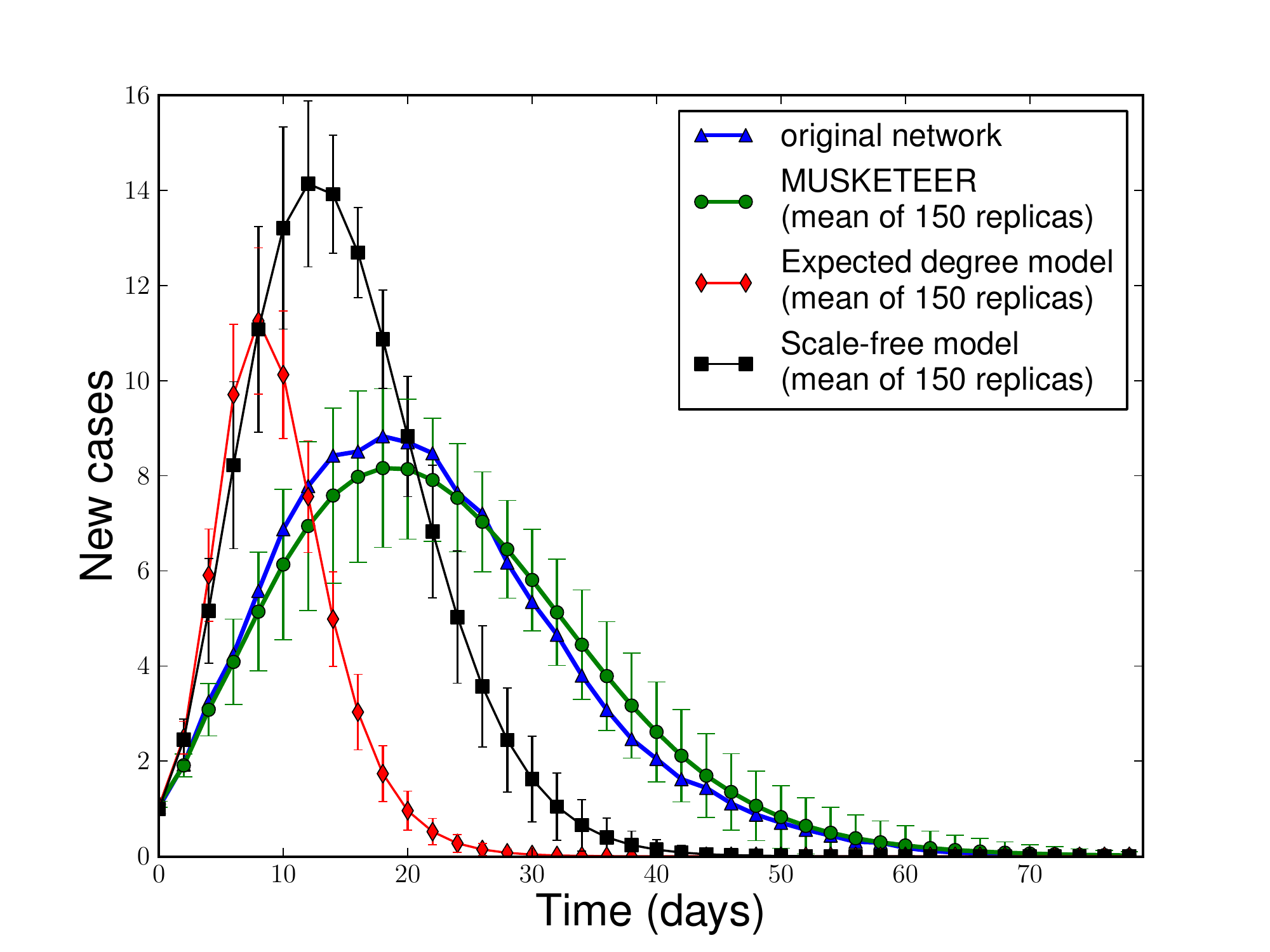}
\end{center}
\caption{SEIR epidemics on the original and replica Colorado Springs sexual contact network.
Graph shows average and one standard deviation (error bars) in incidence on each day of the simulated epidemics, 
across $1000$ runs on the original network and, for each class of synthetic network, $1000$ runs on each of $150$ replicas (a total of $150,000$ runs per class). 
Replica networks were generated using MUSKETEER, the configuration model preserving the original degree sequence~\cite{Chung02}, and the BA scale free model \cite{barabasi} 
(Details in Appendix~\ref{sec:app-alternatives})
\label{fig:SEIR}}
\end{figure}

The MUSKETEER replicas produce epidemiological curves that closely resemble the original network, 
while the other replicas yield outbreaks with much earlier and higher peaks (Figure~\ref{fig:SEIR}).
These results may stem, in part, from the failure of the other models to reproduce
a realistic degree of clustering (see Appendix, Section \ref{sec:app-alternatives}). 
When other aspects of network structure are held constant, clustering has been shown to slow the spread of epidemics via enhanced local connectivity at the expense of global connectivity \cite{volz2011effects}.

MUSKETEER can also be applied to simulate the dynamics of a network evolving through time.
In this mode, the original network is used to generate a replica, $G_1$ which is then used to generate
a replica of a replica, $G_2$, and so on.
When this iterative generation is applied to the Colorado Springs sexual contact network, the network properties appear
to follow a random walk (see Appendix \ref{sec:app-dynamic}). 
A stochastically growing network could also be simulated in this fashion by changing the editing operators.

\section{Discussion}\label{sec:discussion}
MUSKETEER outperforms other widely used network generation models in its ability to introduce unbiased
variability in multiple aspects of network structure and to preserve unspecified properties of
a network that may be critical to the function and dynamics of the system. The latter is accomplished by
randomly editing an existing network rather than creating networks de novo, which retains structure
at multiple levels, including annotation of nodes and edges. However, this also presents an important limitation: the need
for a representative network from which to generate replicas. The real-world fidelity of the replicas hinges on the quality
of the initial network. 

The performance of MUSKETEER also depends on selecting appropriate editing rate at every scale. 
This entails an inherent trade-off between fidelity and variability, which can be explored through trial and error
and should be guided by confidence in the quality of the original network and intended uses of the replicas.
High edit rates and coarse-level editing are advisable when considerable uncertainty exists about 
and/or missing data in the original network. 
The development of robust guidelines for selecting editing rates and methods for generating replicas
from multiple empirical networks are important directions for future work. 

\section{Conclusions}\label{sec:concl}
The MUSKETEER algorithm learns the structure of a network and then perturbs that structure at multiple 
scales to synthesize ensembles of networks with similar properties. In our evaluation, MUSKETEER 
was able to reproduce all local and global network properties considered without a priori specification of those
properties. Thus, it can be applied right out of the box to generate replica networks across diverse domains.  
An implementation of the algorithm is available for free download with Open Source \cite{musketeer:page}.

\par 

\noindent \textbf{Acknowledgments.}
We thank Achi Brandt, Vadas Gintautas, Aric Hagberg, Achim Kempf, Sven Leyffer, Eamon O'Dea, Feng Pan, Oleg Roderick, Dorit Ron, and Cosma Shalizi for encouragement and discussions. 
Jure Leskovec helped with SNAP code and Aaron Clauset, Stephen Q. Muth, and John J. Potterat shared valuable network data.
This work was funded in part by the Department of Energy at the Los Alamos National Laboratory through the LDRD program, and by the Defense Threat Reduction Agency.
The National Institutes of Health have also provided funding through the Models of Infectious Disease Agent Study (MIDAS) program.

\appendix
\section{Technical Details about MUSKETEER} \label{sec:app-technical}
The following paragraphs explain the coarsening and editing steps in more detail.
\subsection{Coarsening and Uncoarsening}\label{sec:coarsening}
The construction of a coarse network $G_{c}=(V_c,E_c)$ from a given $G_f=(V_f,E_f)$ is inspired by the algebraic multigrid (AMG) restriction operators, namely,
\begin{equation}\label{eq:amg}
L_{c}\leftarrow (P_f^c)^T L_f P_f^c,
\end{equation}
where $L_f$ ($L_c$) is the Laplacian corresponding to $G_f$ ($G_c$), and $P_f^c\in \real^{|V_f|\times |V_{c}|}$ describes the connections between fine nodes and corresponding aggregates that form coarse nodes. 
In general, AMG is interpreted as a process of aggregation of the network nodes to define the nodes of the next-coarser network. 
For simplicity, our AMG coarsening scheme assigns each node to a single aggregate. 
Thus, for each row $i$ in $P_f^c$ we allow only one nonzero entry.
This is in contrast to weighted aggregation in which each node can be divided into fractions and different fractions belong to different aggregates.
Construction of (\ref{eq:amg}) is divided into three stages: 
(1) $V_f$ is divided into two disjoint sets: $C$ and $F$;  $C$ are the {\it seeds} of the aggregates, namely, future coarse scale nodes, and $F$ are other nodes that will be aggregated with their neighbors in $C$; 
(2) $P_f^c$ is computed, thereby establishing the assignment of $F$-nodes to their aggregates; and 
(3) couplings (or edges) for $G_c$ are derived from (\ref{eq:amg}).

Various coarsening schemes have been proposed for graphs.
We obtained satisfactory results by following the lines of Ron et al.~\cite{safro:relaxml}.
Briefly, in that algorithm seed nodes are identified based on their high degree and then
from nodes that have no neighbor for a seed.
The seeds form the nodes of the coarsened graph.
Each of the nonseed nodes $u$ is assigned to one of its neighbors that are seeds, based on which neighbor-seed has the highest-weight
edge to $u$.
Two aggregates $A$ and $B$ are connected iff at least one node inside $A$ had a connection to a node inside $B$.
The coarsening process is terminated once the network is reduced to a single node or when its density
(i.e., fraction of edges out of the theoretically possible $\frac{n(n-1)}{2}$) is very high (the default value is $0.9$.)

The high realism of the algorithm and its versatility across domains are due in part to resampling.
Namely, when adding new nodes or edges we copy an existing node or edge, respectively.
For example, to add an edge $\{u,v\}$ to the network at level $i$, we select a random edge of $G_i$, determine how many edges are contained in it,
and insert the same number of edges between the nodes contained in $u$ and those contained in $v$.
In other words, we \emph{resample} the graph when we add features to it.
This process preserves the graph's degree distribution.
Resampling is also used during interpolation when deciding on how many nodes are contained in newly created aggregates 
and how many edges should be created inside an aggregate edge.

The resampling and uncoarsening process are explained in the pseudocode of the interpolation procedure (Alg.~\ref{alg:interpolateM}).
The notation $P_i[u]$ for $u \in V(G_{i+1})$ denotes a subgraph of $G_{i}$ that is deaggregated from coarse node $u$;
Similarly, $P_i[e]$ for $e \in E(G_{i+1})$ denotes the edges of $G_{i}$ that are deaggregated from coarse edge $e$.
\begin{algorithm}[tbh]\label{alg:interpolate}
    \caption{$\textsf{Interpolate}(G_{i+1}, P_i)$}
\begin{algorithmic}[1]\label{alg:interpolateM}
\STATE $NewEdges, NewNodes$ = ListAllNewEdgesAndNodes($G_{i+1}$)

\STATE $G'_{i}  ~ \leftarrow ~ (\emptyset,\emptyset)$ 
\STATE $P'_i   ~ \leftarrow ~ $copy $P_i$

\FOR {$u$ in $NewNodes$}
    \STATE $randomAggregate   ~ \leftarrow ~ P_i$[random node $\in V(G_{i+1})\setminus NewNodes$]
    \STATE $P'_i[u]  ~ \leftarrow ~ $ RelabelNodes[$randomAggregate$]
\ENDFOR

\FOR {$\{u,v\}$ in $NewEdges$}
    \STATE $l  ~ \leftarrow ~ |P_i$[random edge  $\in E(G_{i+1})\setminus NewEdges]|$
    \STATE $U  ~ \leftarrow ~ $  SampleWithReplacement($l$ nodes from $P'_i[u]$)
    \STATE $V  ~ \leftarrow ~ $  SampleWithReplacement($l$ nodes from $P'_i[v]$)
    \STATE $P'_i[\{u,v\}] ~ \leftarrow ~ $ RandomMatching$[U,V]$
\ENDFOR

\FOR {$u$ in $P'_i$}
    \STATE $V(G'_{i}) ~ \leftarrow ~ $ InsertNodes($P'_i[u]$)
\ENDFOR
\FOR {$\{u,v\}$ in $P'_i$}
    \STATE $E(G'_{i}) ~ \leftarrow ~ $ InsertEdges($P'_i[\{u,v\}]$)
\ENDFOR

\STATE \textbf{Return} $G'_i$
\end{algorithmic}
\end{algorithm}

\subsection{Editing}
The function EditEdgesAndNodes edits the edges first, then the nodes.
Editing of edges involves deleting a random number of them, and then inserting approximately the same amount (Alg.~\ref{alg:editM}).
Note that in the pseudocode certain steps are optional (i.e., enabled by the user wishing to tune the algorithm).
The edit rate parameters ($EER_i$ and $NER_i$ for edges and nodes, resp.) are dependent on $i$, the level of coarsening.
\begin{algorithm}[!ht]\label{alg:edit}
    \caption{$\textsf{EditEdgesAndNodes}(G_i)$}
\begin{algorithmic}[1]\label{alg:editM}

\STATE $e_d  ~ \leftarrow ~ $ Binomial($|E(G_i)|, EER_i$)
\STATE $e_a  ~ \leftarrow ~ $ Binomial($|E(G_i)|, EER_i$)
\STATE $n_d  ~ \leftarrow ~ $ Binomial($|V(G_i)|, NER_i$)
\STATE $n_a  ~ \leftarrow ~ $ Binomial($|V(G_i)|, NER_i$)
\STATE $AvgDeg ~ \leftarrow ~ $ AvgDegree($G'_i$)
\WHILE {not deleted $e_d$ edges}
    \STATE $\{u,v\} ~ \leftarrow ~ $ RandomEdge($E(G'_i)$)
    \STATE [Optional] \textbf{if} UniformRandom($[0,1]$) $> AvgDeg$/(Degree($u$)Degree($u$)$)$: \textbf{continue}
    \STATE [Optional] \textbf{if} UniformRandom($[0,1]$) $> 0.5$/MutualNeighbors(u,v): \textbf{continue}
    \STATE $E(G'_i)  ~ \leftarrow ~ E(G'_i) - \{u,v\}$
\ENDWHILE
\STATE [Optional] $E(G'_i)  ~ \leftarrow ~ E(G'_i) \cup \{$Connections to enforce connectedness of the graph$\}$
\WHILE {not added $e_a$ edges}
    \STATE $u ~ \leftarrow ~ $ RandomNode($V(G'_i)$)
    \STATE $d ~ \leftarrow ~ $ SampleInteger($\Prob_{G'_i}$)
    \STATE $Distances ~ \leftarrow ~ $ BreadthFirstSearch($u$ to depth $h$)
    \STATE $v ~ \leftarrow ~ $ RandomNode($Distances$ at distance $d$)
    \IF {$v$ exists}
        \STATE $E(G'_i)  ~ \leftarrow ~ \{u,v\} + E(G'_i)$
    \ENDIF
\ENDWHILE
\WHILE {not added $n_a$ nodes}
    \STATE $anchor ~ \leftarrow ~ $ RandomNode($V(G'_i)$)
    \STATE $source ~ \leftarrow ~ $ RandomNode($V(G'_i)$)
    \STATE $V(G'_i)  ~ \leftarrow ~  u + V(G'_i)$
    \STATE \textbf{if} Degree($source$) = 0: \textbf{continue}
    \STATE $E(G'_i)  ~ \leftarrow ~ \{u,anchor\} + E(G'_i)$
    \WHILE {Degree($u$) $<$ Degree($source$)-1}
        \STATE $Distances ~ \leftarrow ~ $ BreadthFirstSearch($u$ to depth $h$)
        \STATE $v ~ \leftarrow ~ $ RandomNode($Distances$ at distance $d$)
        \IF {$v$ exists}
            \STATE $E(G'_i)  ~ \leftarrow ~ E(G'_i) + \{u,v\}$
        \ENDIF
    \ENDWHILE
\ENDWHILE
\WHILE {not deleted $n_d$ nodes}
    \STATE $V(G'_i)  ~ \leftarrow ~ V(G'_i) - $RandomNode($V(G'_i)$)
\ENDWHILE

\STATE \textbf{Return} $G'_i$
\end{algorithmic}
\end{algorithm}
Note that all the loops have additional checks that exit the loop following a maximal number of iterations.

An optional feature in the function EditEdgesAndNodes avoids changing the degree distribution by using a process
that might be called ``deferential detachment'': 
the probability that an edge $\{u,v\}$ is deleted is inversely related to the product
of the degree of $u$ and the degree of $v$.
As a result, the probability of losing an edge is not dependent on the degree 
(the statement is exact only in graphs with no degree-degree correlations and with all node degrees $> 0$).
The process is analogous to the preferential attachment growth mechanism:
in both cases high-degree nodes either lose fewer edges or gain more than would be the case by chance.
Another optional feature that helps realistically generate graphs with high clustering
is to diminish the probability of deleting $\{u,v\}$ in proportion to the number of mutual neighbors of $u$ and $v$.

Following the deletion process, EditEdgesAndNodes inserts edges by selecting a random attachment point
and selects its new neighbor based on the $\Prob_{G_i}[d]$ distribution.

Turning to nodes, in the first step, it determines the goal number of nodes to be deleted and inserted at that level, a value controlled by a user option for the expected number of edits.
It now deletes nodes through random sampling and inserts the chosen number of nodes.
For each of the new nodes ($u$) it selects a random ``source node'', whose degree it resamples
(note that this process is intentionally simplistic for reasons of computational efficiency and so it ignores dependencies between nodes \cite{Neville08}.)
If the source node has degree $>0$, it proceeds to find an attachment point $v$ (a randomly selected node)
and inserts the edge $\{u,v\}$.
All subsequent edges of $u$ are inserted in the vicinity of the attachment point $v$
based on the distribution $\Prob_{G_i}[d]$.

In is expected that the number of nodes and edges in the edited graph is the same as in the original.
But it is also possible to enlarge or decrease the size of the replica in expectation
by increasing or decreasing the rate of additions of nodes and edges over the rate of deletions.
As a result, a fragment of a network could be used to construct an imprecise copy of the larger network.
This process could also be used to simulate changes in the size of the network over time.

A network of $30000$ edges can be generated in less than a minute even under a very high edit rate (such as $10\%$ over several scales);
With over $110,000$ nodes, a network can be replicated in under 1 minute.
Theoretically, the running time scales as $O(m)$, where $m$ is the number of edges of the original graph.
MUSKETEER currently uses the functionality of the Python programming language and the NetworkX package~\cite{Hagberg08}
both of which trade running time for ease of use.  

MUSKETEER can produce annotated networks, that is, networks that assign attributes to nodes and edges.
Examples of node attributes are age, susceptibility to infection, and history;  
edges might be annotated with strength or relationship type.
MUSKETEER merely preserves the attributes of all nodes and edges that were not edited,
and crudely assigns attributes to new nodes and edges through resampling.
Those correlations between structure and attributes are preserved in unedited components.

\section{Extended Data on the Performance of MUSKETEER}\label{sec:app-extended-perf}
We present the results of evaluation of the MUSKETEER network generator on several real-world networks (see Table \ref{tab:networks}) of different types by using two sets of parameters. 
In the first set (P1) the networks are edited at the first two fine levels: 8\% at the finest level and 7\% at the next coarser level, for both nodes and edges. 
In the second set (P2) the networks are edited at 5 levels: 5\% (finest), 4\%, ... , 1\%, 1\%, ..., 1\% (coarsest), again in both nodes and edges.
For each network, $50$ replicas were generated for each set of parameters, and then several properties of comparison were measured. 
Note that here the measure of degree of assortativity is Newman correlation coefficient \cite{Newman03mixing}.
When the original data included more than one connected component, only the biggest connected component was considered, in order to simplify the comparison.
Note also that some metrics of comparison are meaningless (have no practical relevance) for several networks and thus are not presented for them.\\
{\bf Yeast (protein-protein interactions network).} This network was analyzed in \cite{yeastnet} in order to uncover hidden topological structures that may consist of biologically relevant functional information. 
One of the main results of that work was understanding the structure and amount of quasi-cliques and quasi-bipartites. 
Being NP-hard to detect, such properties have been approximated and analyzed by using tools from spectral graph theory. 
These notions are strongly connected to metrics such as clustering, modularity, and eigenvector centrality. 
These metrics are preserved in the generated networks (see Table \ref{tab:networks}). 
Average all-pairs distance (by shortest path), the harmonic average distance, and betweenness centrality were changed by at most 4\% each.\\
{\bf Electrical power grid of the western United States.} One of the most interesting properties of many real-world networks is the ``small-world'' property. 
Watts and Strogatz \cite{Watts1998} demonstrated that this network  has this property; thus, clustering, modularity, and shortest paths-related statistics are of particular importance. 
Note that in contrast to clustering and modularity that are preserved correctly, the average distance has been decreased. 
The reason is that at the several critical ``short cut'' nodes the editing process has created additional short cuts that influence the average value (an example of the small-world effect \cite{Watts1998}). 
This can be easily corrected by a tuning parameter available in MUSKETEER, which blocks insertion of long-range edges.  Importantly, the small-world property has not been corrupted.\\
{\bf Krebs' network.} The terrorist cell network belongs to the type of data that is extremely difficult to obtain and use \cite{Krebs02} because of such factors as incompleteness, 
dynamics, and fuzzy boundaries. 
Different types of cuts, clustering, and centrality measures are among the key analytical tools for understanding the behavior of a terrorist cell. 
We observe that all metrics are well preserved in both parameter settings.\\
{\bf English words transformation.} This network represents the relationships between words that are within Levenshtein distance 1 from each other \cite{na-texts}. 
Applications of analysis of such networks arise in linguistics. 
Only the biggest and most informative connected component was replicated in cases where the other components were incomparably smaller. 
Both clustering and modularity are well preserved whereas the average distance metrics are decreased because of the same reason as for the power grid network.

\begin{table}[H]
\caption{Networks used in evaluating MUSKETEER, and the realism of generated networks.  Definitions are given in the main text.
}
\label{tab:networks}
\footnotesize
\centering
\begin{tabular}{|l|cccc|}
\hline
Graph & Metric & Original & P1 & P2 \\
\hline
\multirow{11}{*}{Yeast protein-protein interactions} & nodes & 2224 & 2225.14 & 2215.28\\
& edges & 6609 & 7091 & 6797.12 \\
& clustering & 0.14 & 0.14 & 0.14\\
& modularity & 0.59 & 0.53 & 0.56  \\
& avg between. central.	& 0.00	& 0.00	& 0.00\\
& eigenvector centrality & 0.01 & 0.01 & 0.01\\
& mean ecc & 7.61 & 7.96 & 8.54\\
& avg distance	& 4.44	& 4.22	& 4.34\\
& harmonic avg distance & 0.00 & 0.00 & 0.00\\
& avg flow closeness & 0.92 & 1.16 & 1.11 \\
& powerlaw exp & 2.26 & 2.35 & 2.37\\
& degree assortativity & -0.11 & -0.07 & -0.07 \\
\hline
\multirow{11}{*}{Power grid} & avg nodes & 4941 & 4950.10 & 4768.74\\ 
 & avg edges & 6594 & 6674.68 & 6356.60\\
 & clustering & 0.08 & 0.08 & 0.08\\ 
 & modularity & 0.93 & 0.91 & 0.92\\ 
 & betweenness centrality & 0.00 & 0.00 & 0.00\\
 & eigenvector centrality & 0.00 & 0.00 & 0.00\\
 & mean ecc & 34.54 & 25.45 & 26.95\\
 & avg distance & 19.029 & 13.26 & 13.94\\ 
 & harmonic avg distance & 0.00 & 0.00 & 0.00 \\
 & avg flow closeness & 0.20 & 0.29 & 0.27\\
 & power law exp & 4.48 & 4.65 & 4.68\\
 & degree assortativity & 0.00 & 0.00 & -0.01 \\
\hline
\multirow{11}{*}{Krebs' network} & avg nodes & 62 & 62.24 & 61.74\\ 
 & avg edges & 152 & 152.88 & 157.9\\ 
 & clustering & 0.49 & 0.44 & 0.44 \\ 
 & modularity & 0.53 & 0.52 & 0.50\\ 
 & betweenness centrality & 0.03 & 0.04 & 0.03\\  
 & eigenvector centrality & 0.09 & 0.09 & 0.09 \\
 & mean ecc & 4.39 & 5.72 & 5.19\\
 & avg distance & 2.90 & 3.19 & 2.97\\ 
 & harmonic avg distance & 0.04 & 0.04 & 0.04\\ 
 & flow closeness & 0.99 & 0.95 & 1.07\\   
 & power law exp & 3.18 & 3.44 & 3.34\\ 
  & degree assortativity & -0.08 & -0.06 & -0.07 \\
\hline
\multirow{11}{*}{Lederberg citations} & avg nodes & 8212 & 8308.40 & 8259.06\\ 
 & avg edges & 41430 & 43454.89 & 40180.08 \\ 
 & clustering & 0.32 & 0.28 & 0.29\\ 
 & modularity & 0.70 & 0.61 & 0.64\\ 
 & betweenness centrality & 0.00 & 0.00 & 0.00 \\
 & eigenvector centrality & 0.00 & 0.01 & 0.01\\
 & mean ecc & 11.62 & 8.41 & 9.17\\
 & avg distance & 4.41 & 3.97 & 4.17\\ 
 & harmonic avg distance & 0.00 & 0.00 & 0.00\\
 & flow closeness & 1.55 & 1.70 & 1.56\\ 
 & power law exp & 2.18 & 2.17 & 2.20\\ 
 & degree assortativity & -0.10 & -0.09 & -0.10\\ 
\hline
\multirow{8}{*}{English words transformation} & avg nodes & 24831 & 24801.78 & 25178.56\\ 
& avg edges & 71014 & 75022.11 & 73868.33\\
& clustering & 0.24 & 0.25 & 0.24 \\
 & modularity & 0.83 & 0.78 & 0.77 \\ 
 & betweenness centrality & 0.00 & 0.00 & 0.00\\
 & avg distance & 9.26 & 7.224 & 7.05\\
 & harmonic avg distance & 0.00 & 0.00 & 0.00\\ 
 & power law exp & 2.80 & 2.73 & 2.81\\
 & degree assortativity & 0.61 & 0.37 & 0.30\\ 
\hline 
\end{tabular}
\end{table}

\normalsize

\section{Dynamic Networks with MUSKETEER}\label{sec:app-dynamic}

\begin{figure}[!bth] \begin{center}
\includegraphics[width=0.7\textwidth]{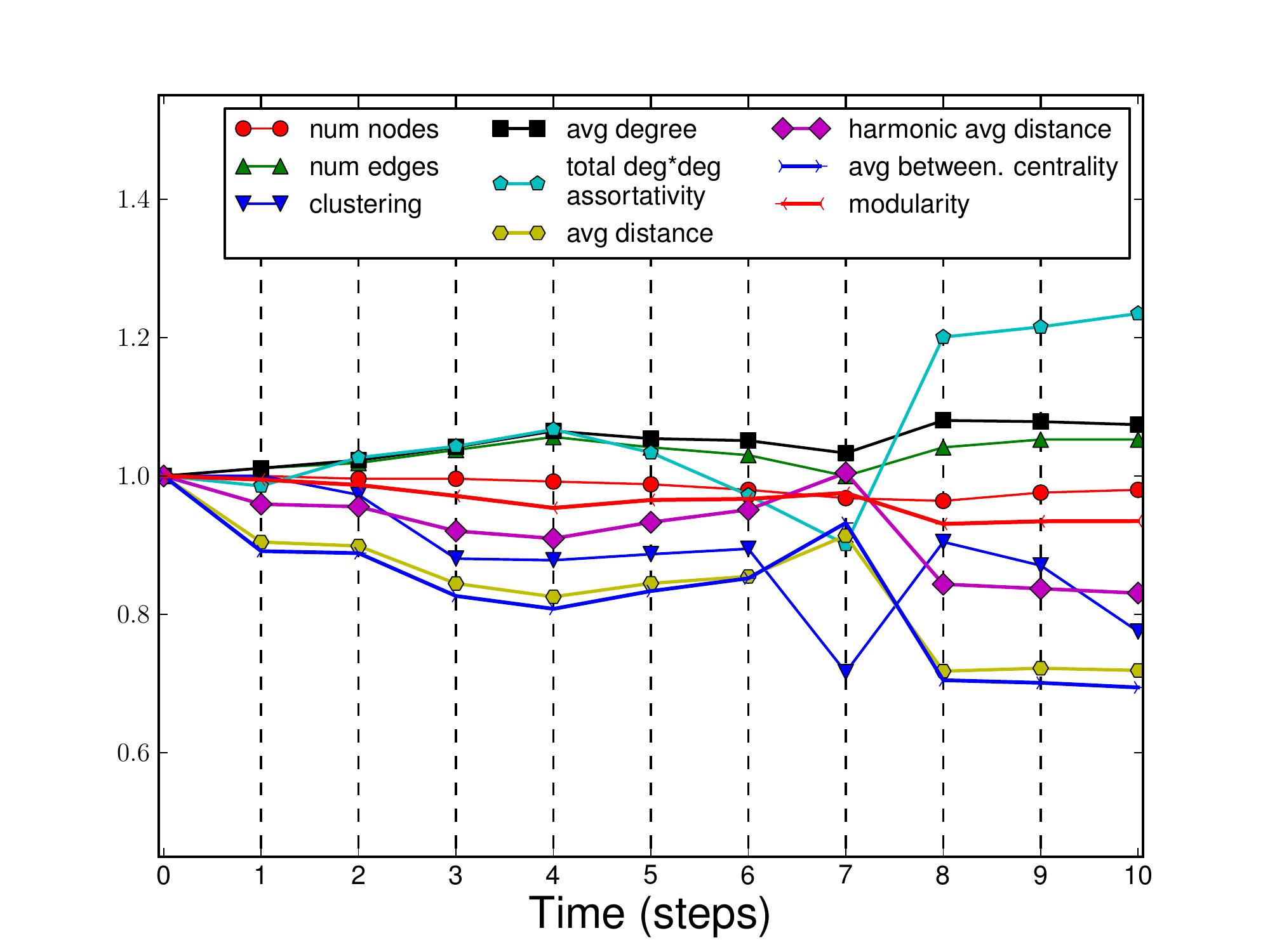}
\end{center}
\caption{Using MUSKETEER to simulate a dynamic network.
At time $0$ the simulation starts at the Colorado Springs network \cite{Potterat02},
which is then evolved in time using 10 rounds of iterated replications.
At every replication, the editing rate is 1/10 of the ones used in the previous simulations.
Observe that the properties exhibit effectively a random walk, as desired if we wish to simulate network dynamics.
For every reported property, the vertical scale is normalized by the value of the original graph.
\label{fig:snapshots}}
\end{figure}

Because the multiscale generator preserves properties of the original data, it can be used to simulate dynamics not only on the network
but also of the network.
Namely, a generated replica can represent the topology of the network as it might exist following a short period of natural change.
The replica could itself be replicated, giving a slightly more extended forecast, and so on, going further and further in time.
If the network is not expected to grow or shrink, the structure of the network is expected to follow a random walk.
Figure ~\ref{fig:snapshots} shows that indeed it does, suggesting the evolution is simulated realistically.
One can actually use MUSKETEER to simulate networks that grow over time, by setting the editing process
to insert more nodes than it deletes.
We found that this approach preserves not only most properties of the graph but also its original geometry.
For example, if one takes a regular square grid and increases the number of nodes by a factor of 2, the resulting network is still strongly square grid-like, 
as detected by solving systems of spring forces (often used by visualization algorithms). 
See illustrations in Appendix \ref{sec:app-illustrations}.

\section{Performance of Several Existing Network Generators}\label{sec:app-alternatives}
In the introduction we mentioned a number of existing approaches to modeling and generation synthetic graphs.
In this section we review them and evaluate their realism and entropy based on the same criteria as needed for MUSKETEER.

\subsection{Overview of Existing Methods}
Many existing network generators construct a random graphs with specified properties
(see surveys in \cite{chakrabarti2006graph,brase2009modeling,dunlavy2009mathematical}).
Some generators take as input a number of properties of the original network and then construct a random network that exhibits those properties.
For instance, many generators reproduce the degree distribution \cite{Albert2002,Newman03thestructure,mahadevan2006systematic}, clustering \cite{Bansal09}, and the number of small subgraphs \cite{robins2007introduction}.
Those generators are attractive because they frequently produce networks with well-characterized structural properties \cite{erdos1960erg}.
Another kind of generator imitates the hypothesized growth of the original network \cite{barabasi,krapivsky,Leskovec08}.
Some models incorporate geographic information 
\cite{Watts1998} or focus on particular domains such as cyber networks \cite{medina,dunlavy2009mathematical}.

Another common approach for network generation is based on edge swapping.
In this method one starts with a network and edits the edge set in a process that provably maintains the degree distribution.
This involves randomly selecting a pair of edges $\{x_1,x_2\}$ and $\{y_1,y_2\}$ and replacing
them with $\{x_1,y_1\}$ and $\{x_2,y_2\}$.
An important characteristic of this method is that it defines a Markov chain that is guaranteed
to eventually uniformly sample from all the networks with this degree sequence \cite{mihail2003markov} 
and could even maintain other selected properties of the original network.
Our approach is similar in that we modify an existing network. However, we introduce a number
of new stochastic editing processes such as resampling. Moreover, in our approach changes at fine scales avoid insertion of nonlocal edges; and these local changes are performed at multiple scales, thus adding the ability to control the similarity with the original network across the scales.

A different random construction strategy is based on Kronecker graphs \cite{kronmodel} (see also related works \cite{chakrabarti2006graph,palla2010multifractal}).
Here the graph is generated from a small square matrix that is used for iterative growth of the network.
Such generation processes are plausible because many real networks are organized hierarchically \cite{barabasi2003sfa, palla2005uncovering}.
Because of the small number of parameters, however, such processes are unlikely to be able to generate the realistic network with a hidden structure, and indeed they have well-known limitations \cite{seshadhri2011depth}.
Another limitation is the implicit assumption that the network is self-similar across different scales of distance and time.
Actually, substantial evidence shows that many complex systems are self-dissimilar, that is, organized differently across scales (see \cite{Carlson02, Itzkovitz05, palla2005uncovering, Wolpert07, Binder08, Mones12}).
Thus, a network may by formed through different processes at different scales and is constrained by different laws at each scale.
For example, connections in a supply network (from factories to wholesalers to retailers to consumers) are determined in its large scale by costs of long--range transportation and by social and competitive forces at the small scale.
In order to capture those differences, our strategy learns the structure of the network at each of its different scales.

\subsection{Performance of Selected Existing Methods}
We selected seven methods among the most prominent, commonly used, and versatile options.
We hope to extend this list as some of the software for the generators is published.
\begin{enumerate}
\item An  Erd\H os-R\'enyi random graph that matches the size $n$ and edge density of the original network~\cite{erdos1960erg}.
\item A small world graph with 4 nearby neighbors per node and random edges with probability equal to
the density of the original network~\cite{Watts1998}.
\item A scale-free graph~\cite{barabasi} with $n$ and $m$ adjusted to match the size and density of the original network.
\item A random graph with size and expected degree distribution that matches the original~\cite{Chung02}.
\item A graph constructed by randomly rewiring $30\%$ of the edges (the same edit rate
as when evaluating MUSKETEER in Fig.~\ref{fig:edits}).
\item A graph constructed by swapping $30\%$ of the edges while preserving the degree distribution and connectivity~\cite{Gkantsidis03}.
\item A Kronecker random graph generated by using the SNAP library \cite{kronmodel} by fitting a 2x2 matrix (using the default 50 restarts)
and then generating a graph to 256 nodes (the closest power of 2 to the original graph).
The direction of the edges was ignored.
\end{enumerate}
Note that here Erd\H os-R\'enyi and scale-free random graphs are used as generators, as compared with  Fig.~\ref{fig:parallel-2models} where
they were used as input to MUSKETEER.

Before we present the performance of those algorithms, we grant that our evaluation makes arbitrary choices for the input network (\cite{Potterat02} Fig.~2A) and
 makes no effort to tune the alternative generators, instead running them with a default parameter set.
A more comprehensive evaluation is outside the scope of this work,
but the results are unlikely to be very different because they are consistent with some of the known properties of those generative models.

\begin{figure}[htbp]
\centering
\subfloat[MUSKETEER]{\label{fig:MUSKETEER-gen}\includegraphics[width=0.35\textwidth]{replica_vs_original_potterat__1__2012_05_17__23_06_26}}
\subfloat[Erd\H os-R\'enyi Random Graph]{\label{fig:er-gen}\includegraphics[width=0.35\textwidth]{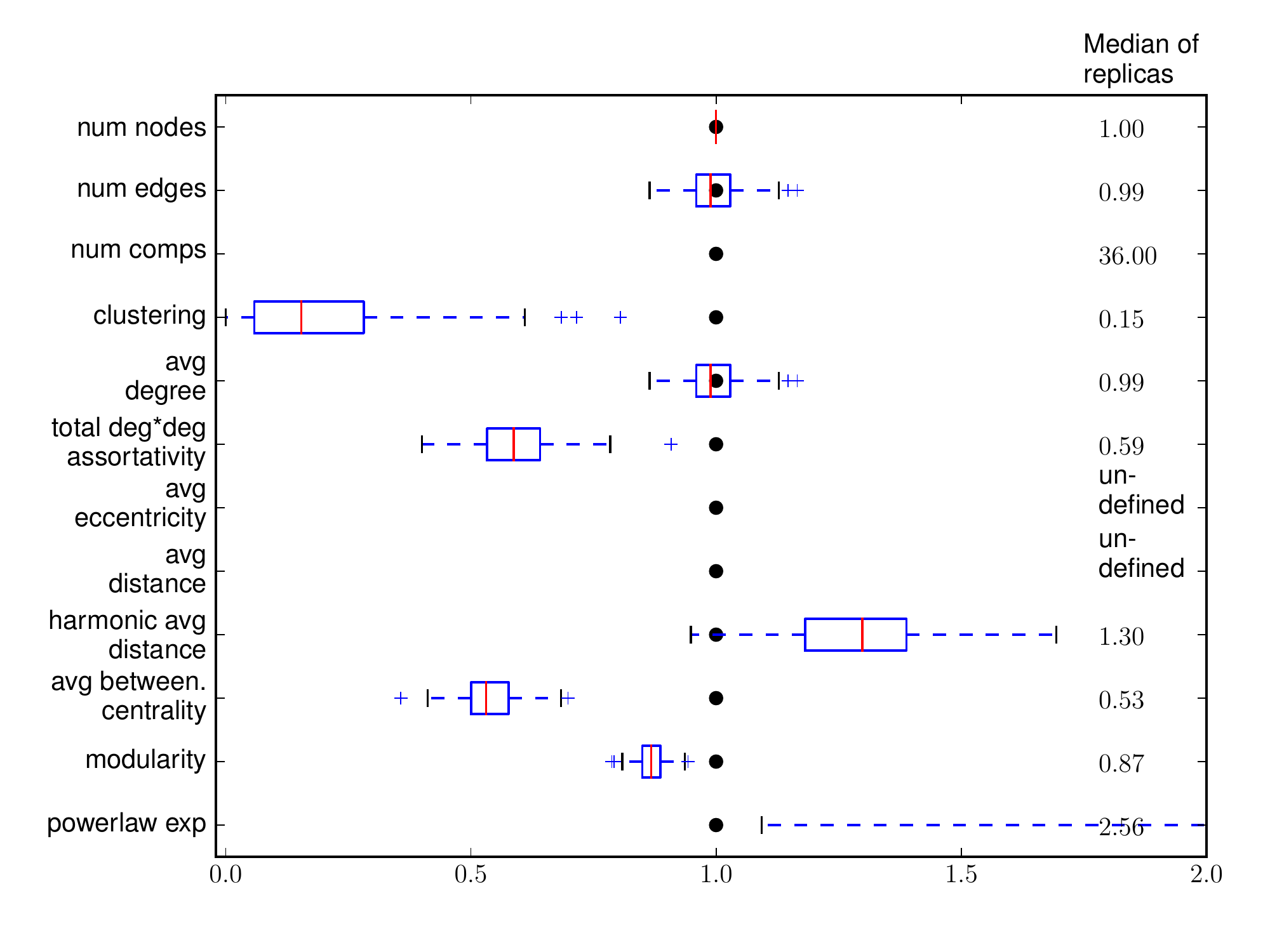}}\\
\subfloat[Scale-Free Graph]{\label{fig:scalefree}\includegraphics[width=0.35\textwidth]{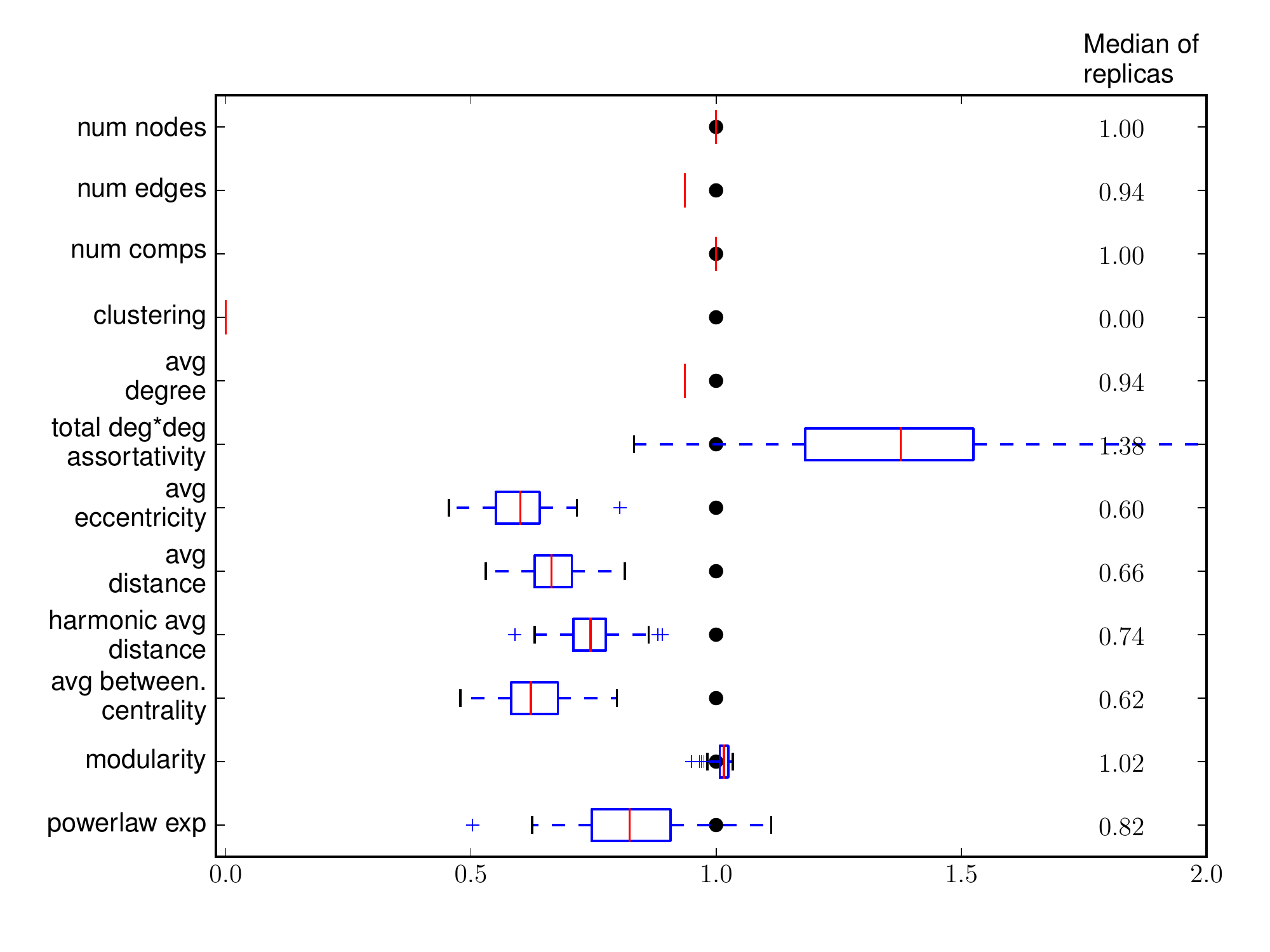}}
\subfloat[Small World Network]{\label{fig:sw}\includegraphics[width=0.35\textwidth]{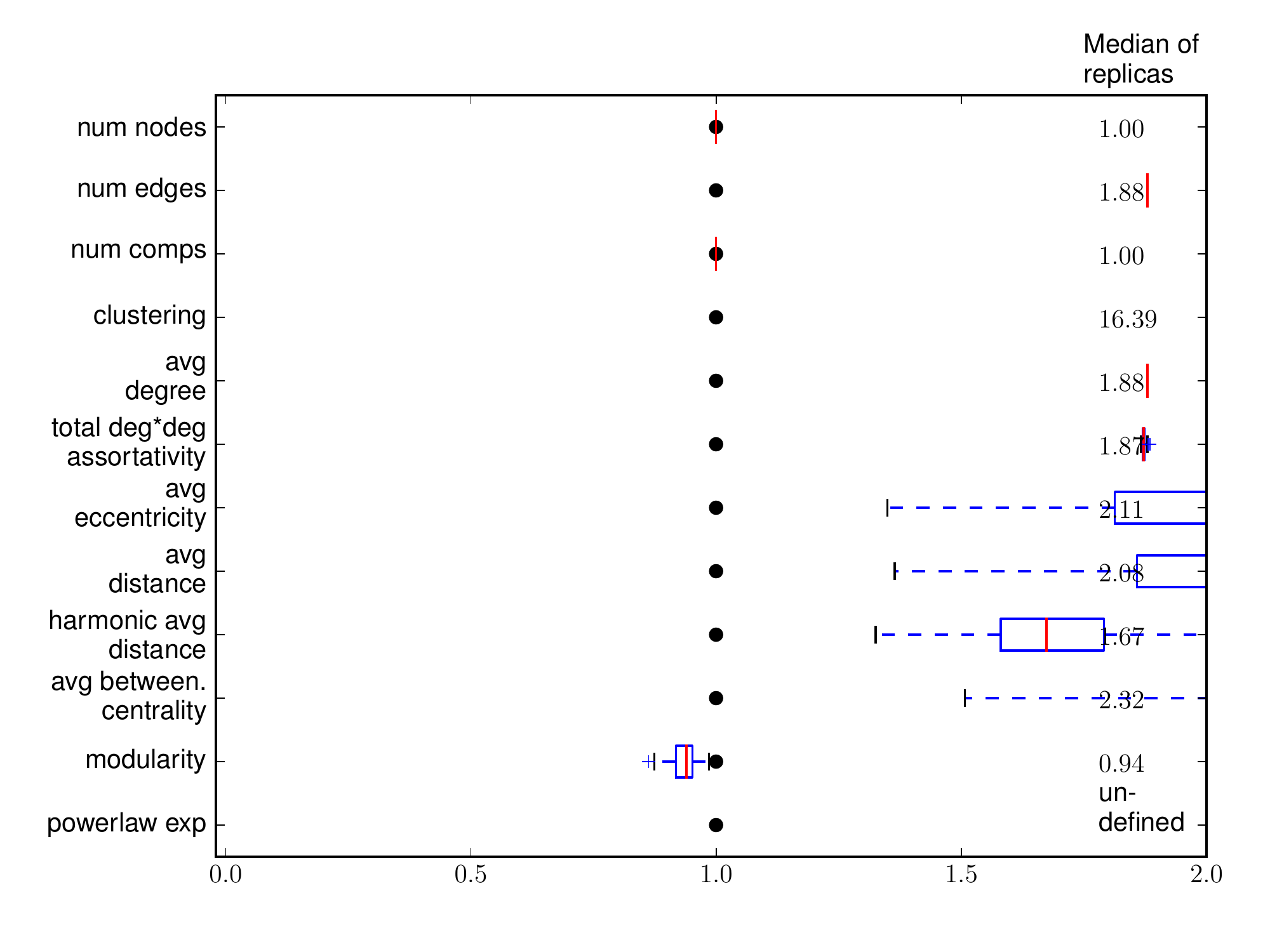}}\\
\subfloat[Expected Degree]{\label{fig:ed}\includegraphics[width=0.35\textwidth]{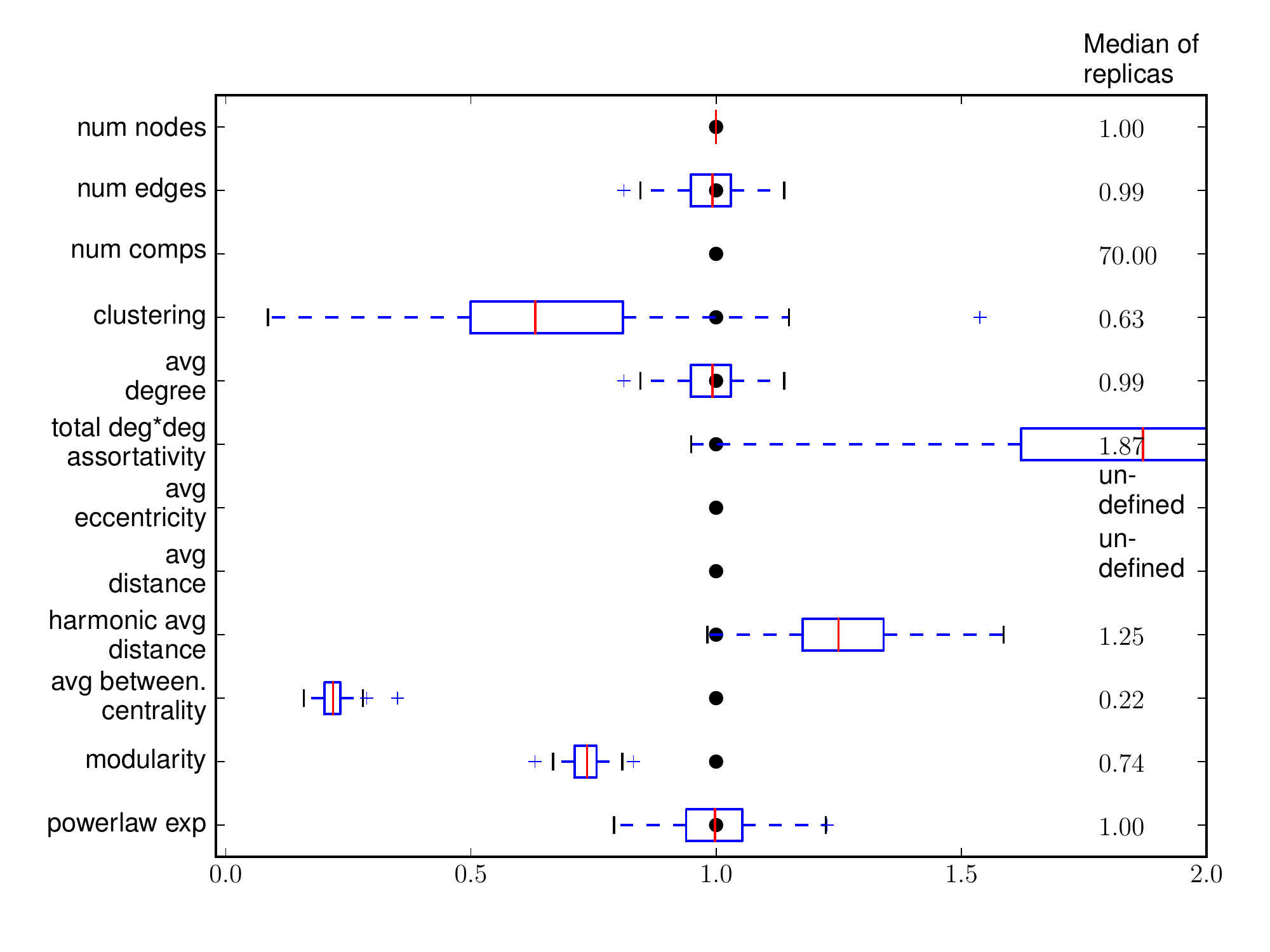}}
\subfloat[Kronecker Random Graph]{\label{fig:kr}\includegraphics[width=0.35\textwidth]{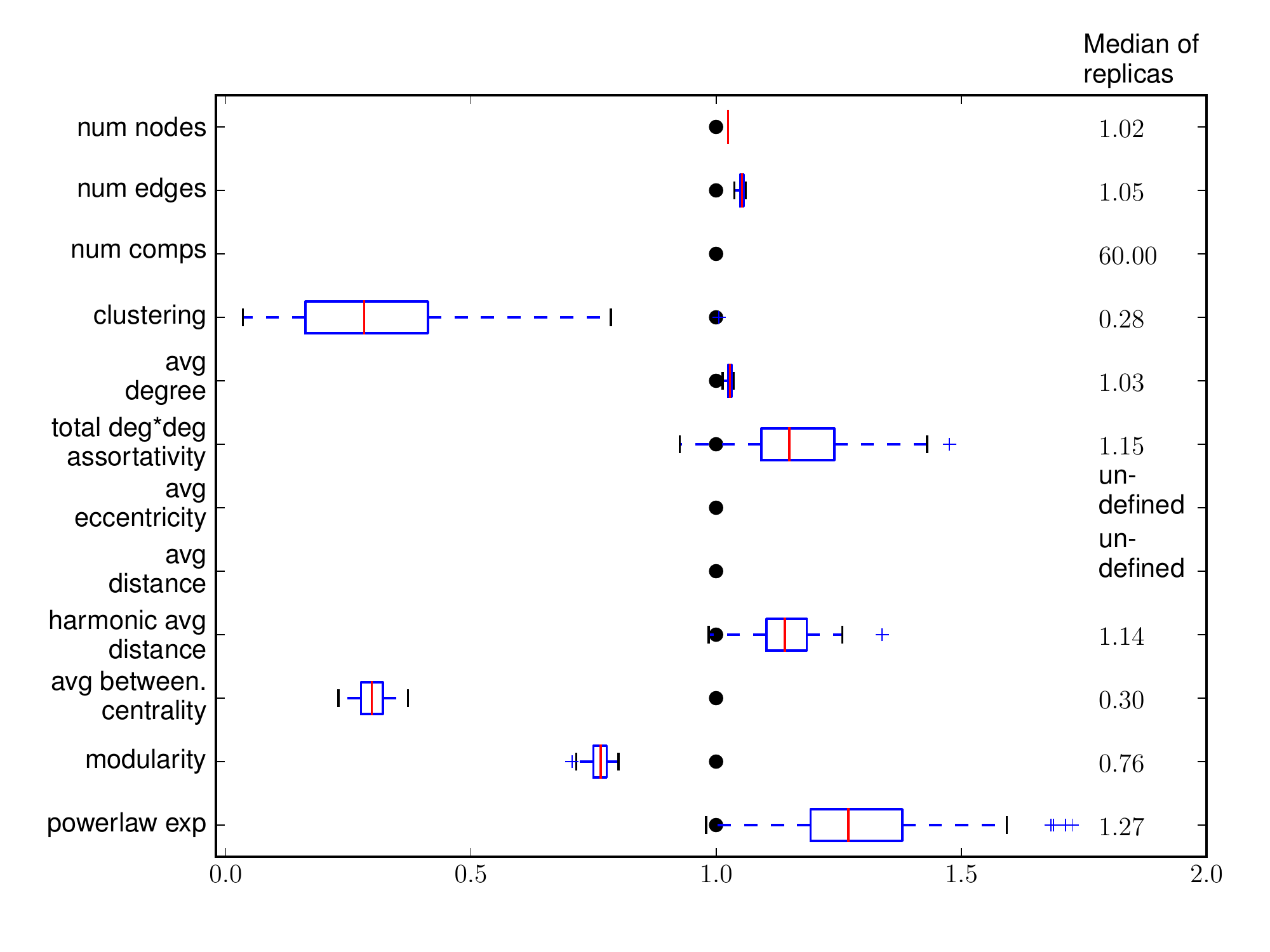}}\\
\subfloat[Random Edge Edit]{\label{fig:re}\includegraphics[width=0.35\textwidth]{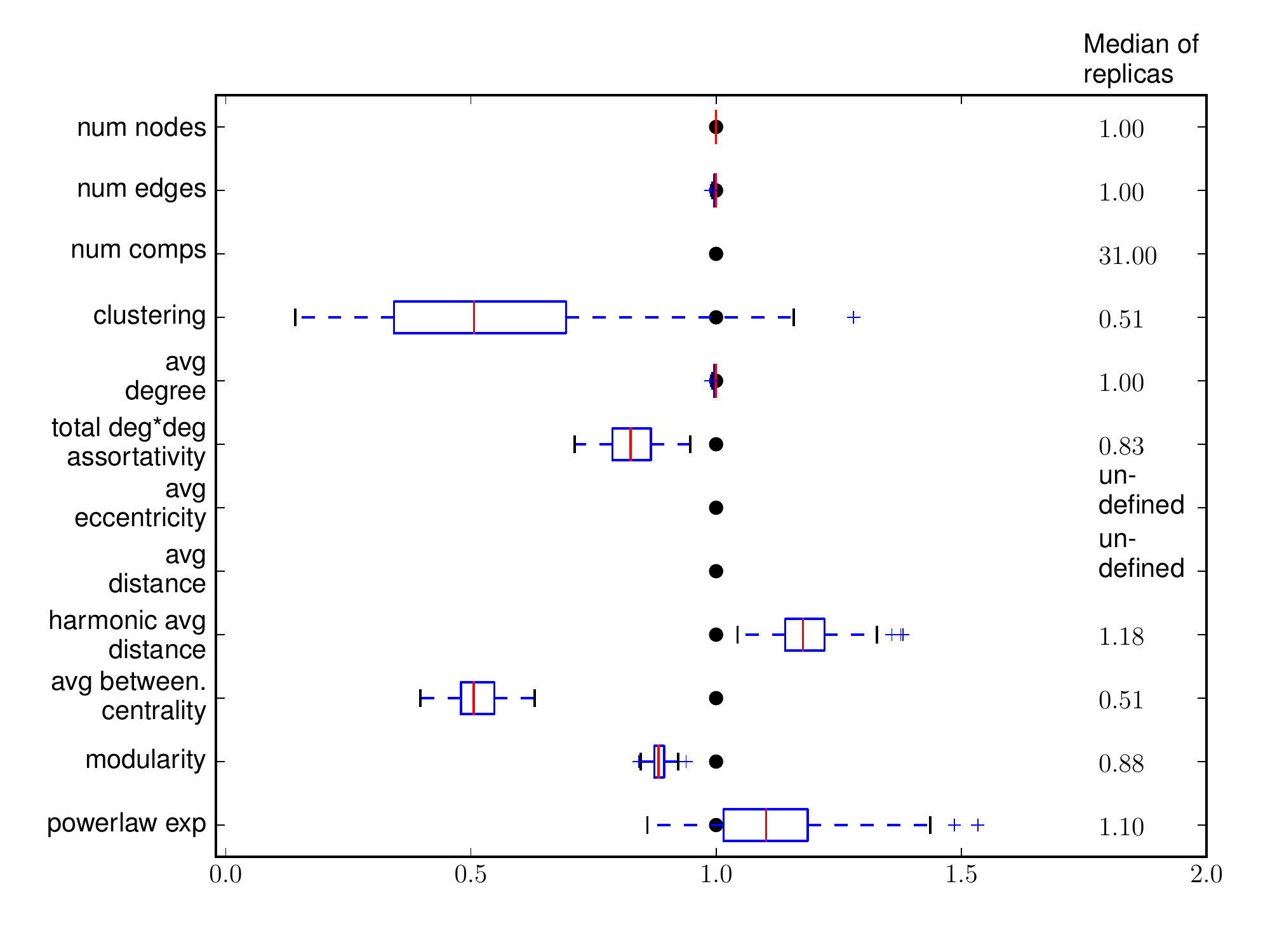}}
\subfloat[Random Edge Swap]{\label{fig:rs}\includegraphics[width=0.35\textwidth]{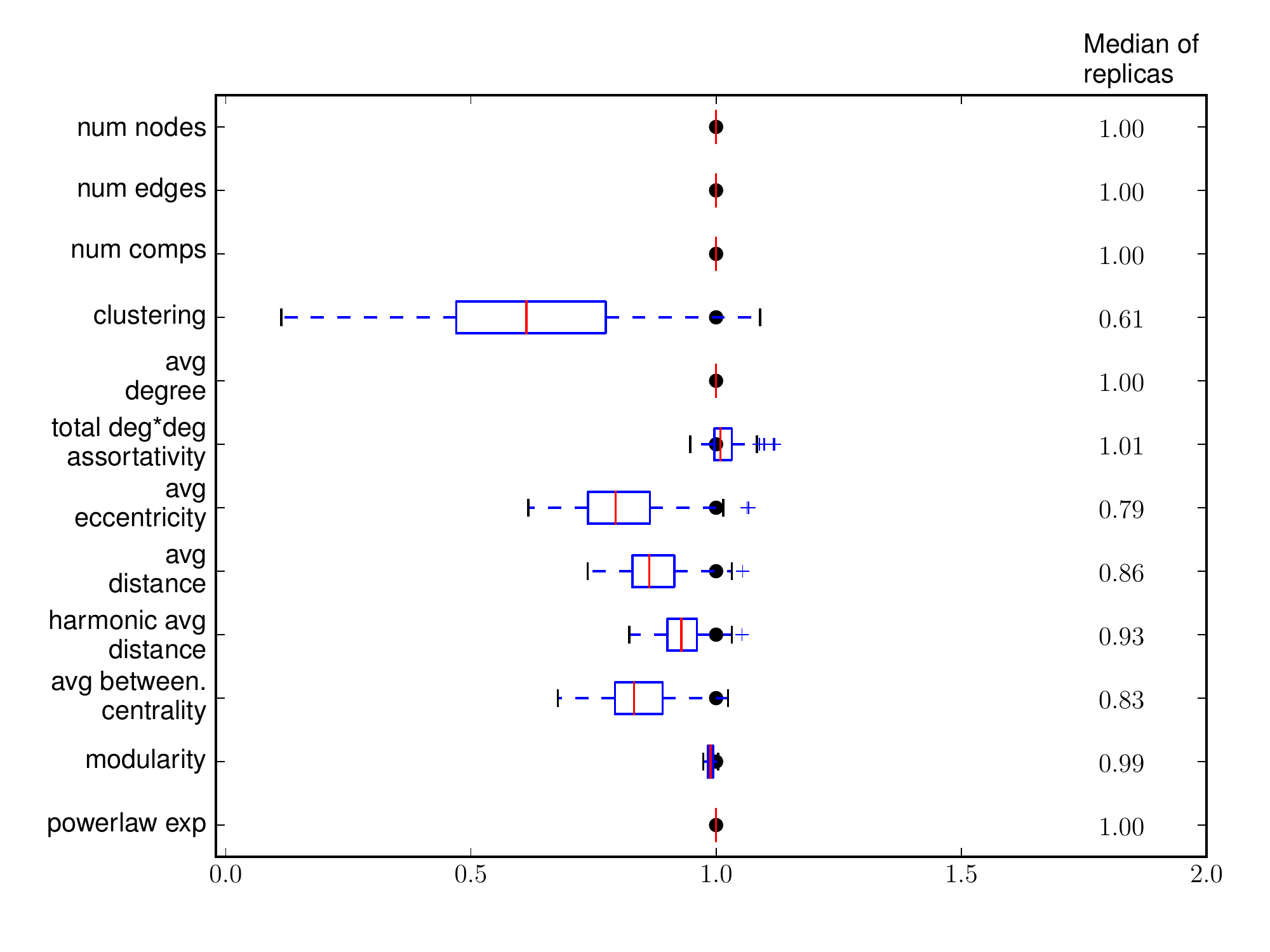}}
\caption{Realism of 8 network generators (Part II, box plots) with respect to the original network (black dot at $1.0$)\label{fig:competitors}.
(a) reproduces Fig.\ref{fig:potterat} from above, for convenience of comparison.
Observe that some metrics were undefined, as when the generated network is not connected, or are off the scale.
For example, in the Small World model, the clustering is $0.49-0.50$,
which is much greater than the clustering of the original graph ($0.03$).
The 7 existing generators exhibit consistent biases along many of the metrics and extremely low entropy in others.}
\end{figure}

The simulations were run as in the original evaluation of MUSKETEER (Fig.~\ref{fig:edits}):
150 replicas of the original graph were generated by using each of the algorithms
and then compared in their properties with the original graph.
The results are shown in Fig.~\ref{fig:competitors}.
We find in those simulations that MUSKETEER gives substantially higher realism in the generated networks.

This high realism is not due to insufficient editing of the original network;
indeed, MUSKETEER generates networks that are substantially different from the original network,
with clustering changed by more than $10\%$ for over half the replicas, to mention one metric.
It is able to do this while avoiding any systematic bias in most of the measures of realism.
The biases we observed in the existing generators are consistent with their known properties,
such as low clustering in the random-graph approach \cite{Newman03thestructure}, and so can be expected to be observed
with other input networks.
Those biases might be partially remedied in some cases by post-processing (e.g., connecting disconnected components);
but with most properties, such as the aforementioned low clustering, such an approach is not readily possible.

An important limitation of this analysis is that we focus on the realism of the replicas.
In many applications the running time or the space requirements might also be important.
While our algorithm is linear in the number of edges, its performance has a larger constant
factor than a model such as the Erd\H os-R\'enyi random graph.
Also, many of the competing network generators have their own advantages, for example a well-developed theory or relative simplicity.
Frequently, they were proposed as stylized theoretical constructs (toy models)
to explain phenomena such as the Small World \cite{Watts1998}, rather than as realistic models of real networks.
The multiscale approach of the MUSKETEER algorithm is designed to achieve the opposite objective:
to construct highly realistic synthetic networks.

\section{Examples of Generated Networks}\label{sec:app-illustrations}

\begin{figure}[H]
\centering
\subfloat[Original network]{\label{fig:orig1}\includegraphics[width=0.35\textwidth]{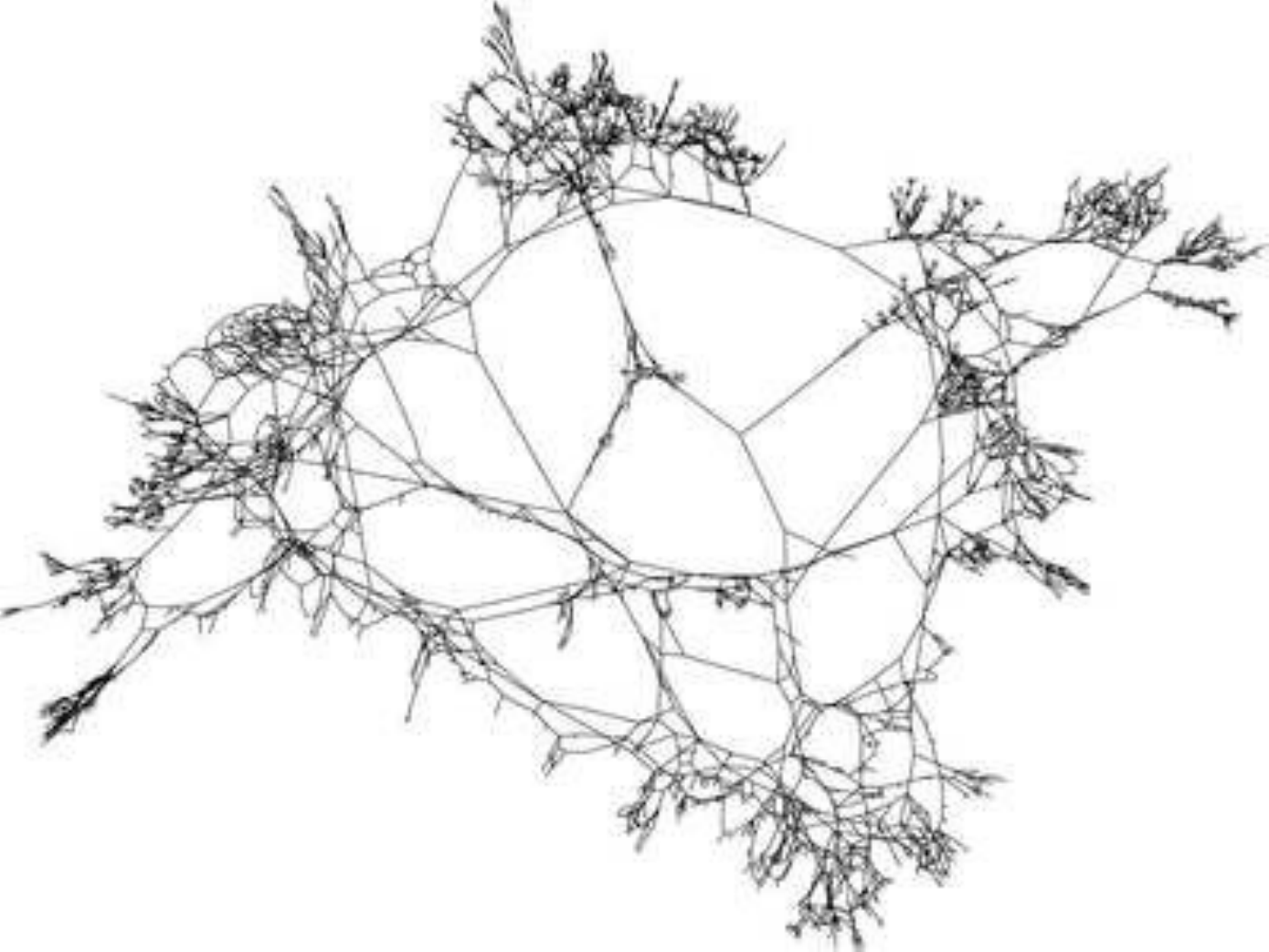}}
\subfloat[Fine level edge edits]{\label{fig:fine1}\includegraphics[width=0.35\textwidth]{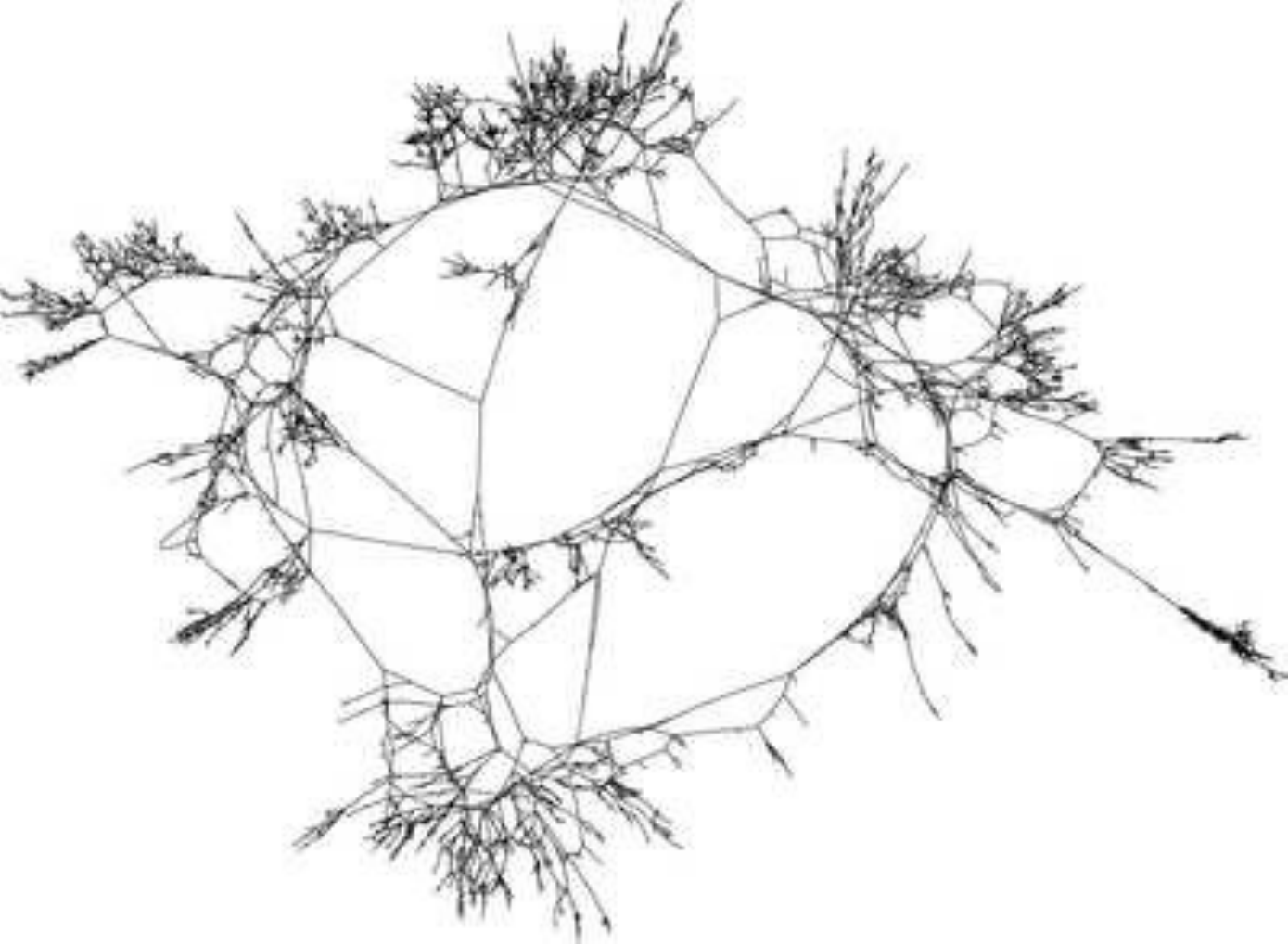}}\\
\subfloat[Small number of coarse edge edits]{\label{fig:coarse1}\includegraphics[width=0.35\textwidth]{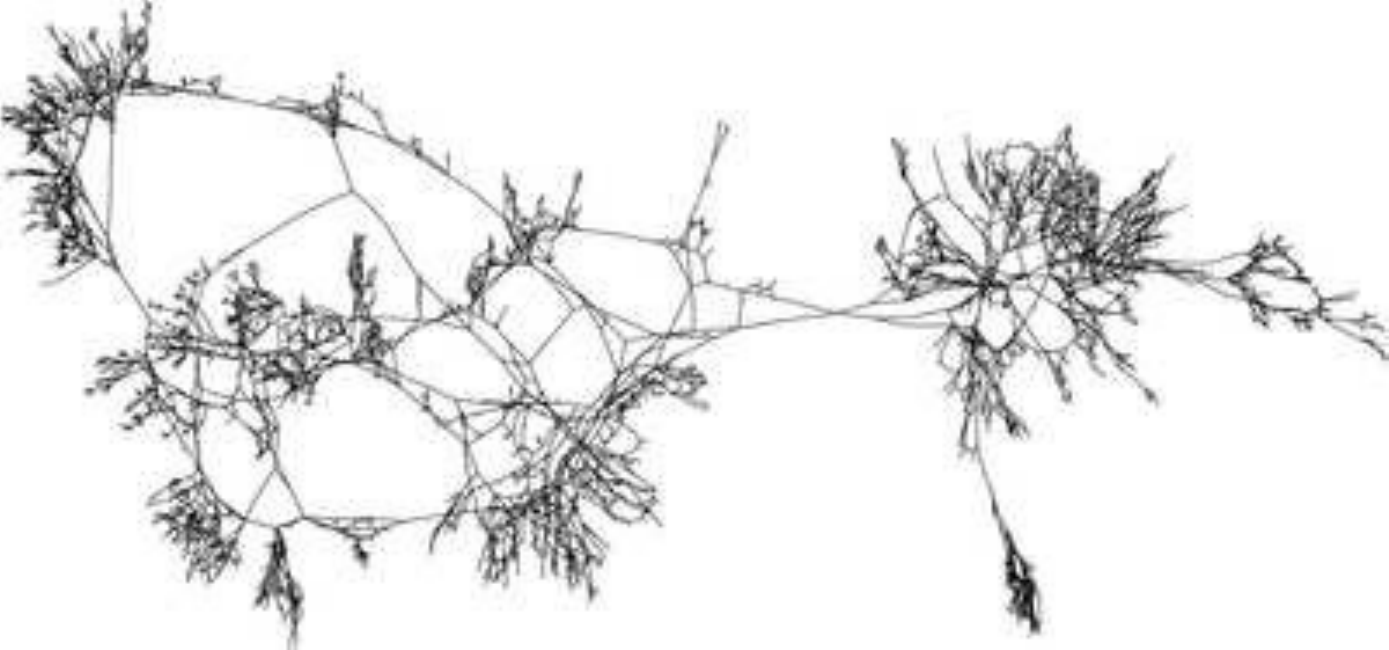}}
\subfloat[Fine edge edits with node growth]{\label{fig:finegrowth1}\includegraphics[width=0.35\textwidth]{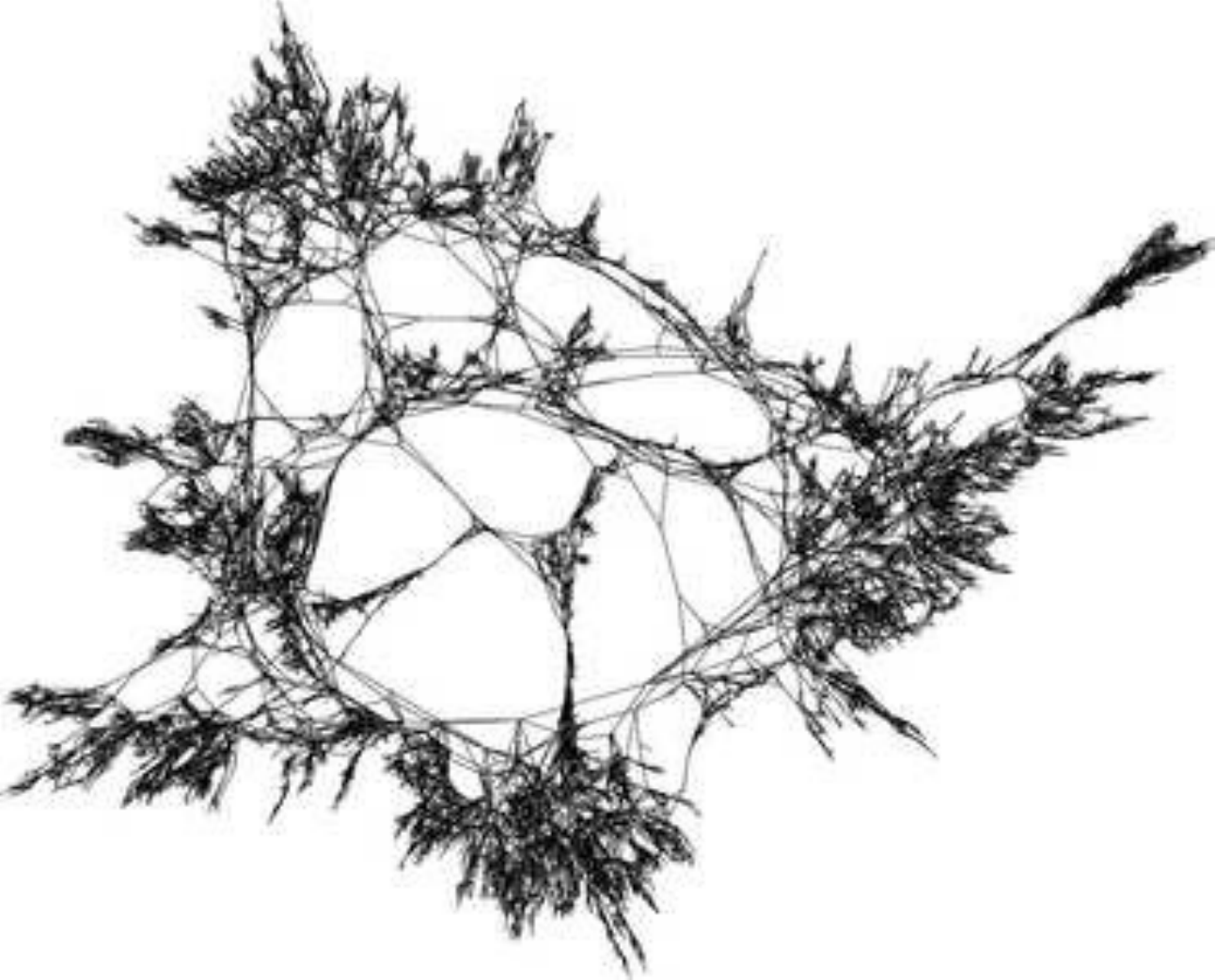}}\\
\subfloat[Several coarse edge edits]{\label{fig:globchanges1}\includegraphics[width=0.35\textwidth]{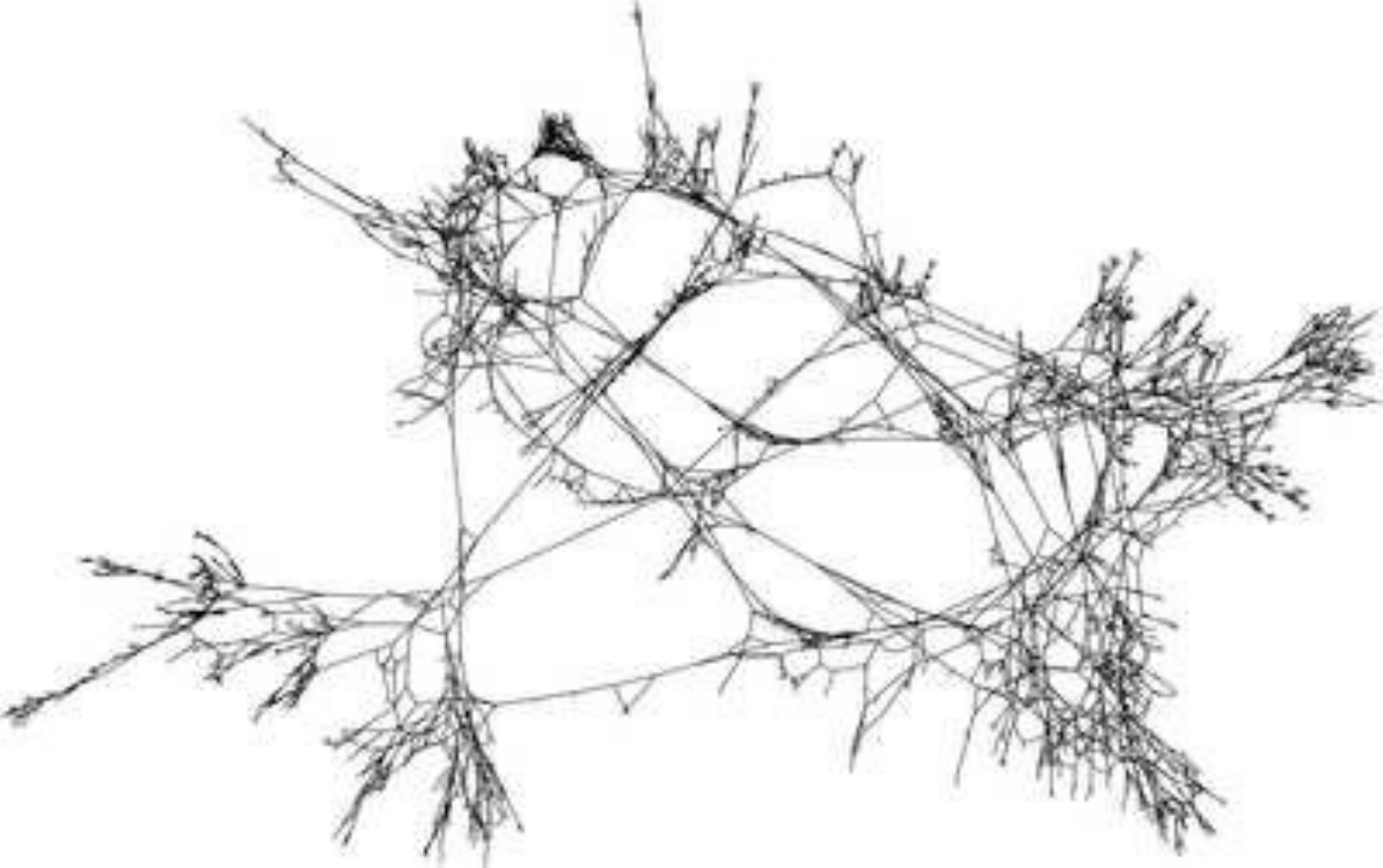}}
\subfloat[Coarse edge edits and node growth]{\label{fig:globgrowth1}\includegraphics[width=0.35\textwidth]{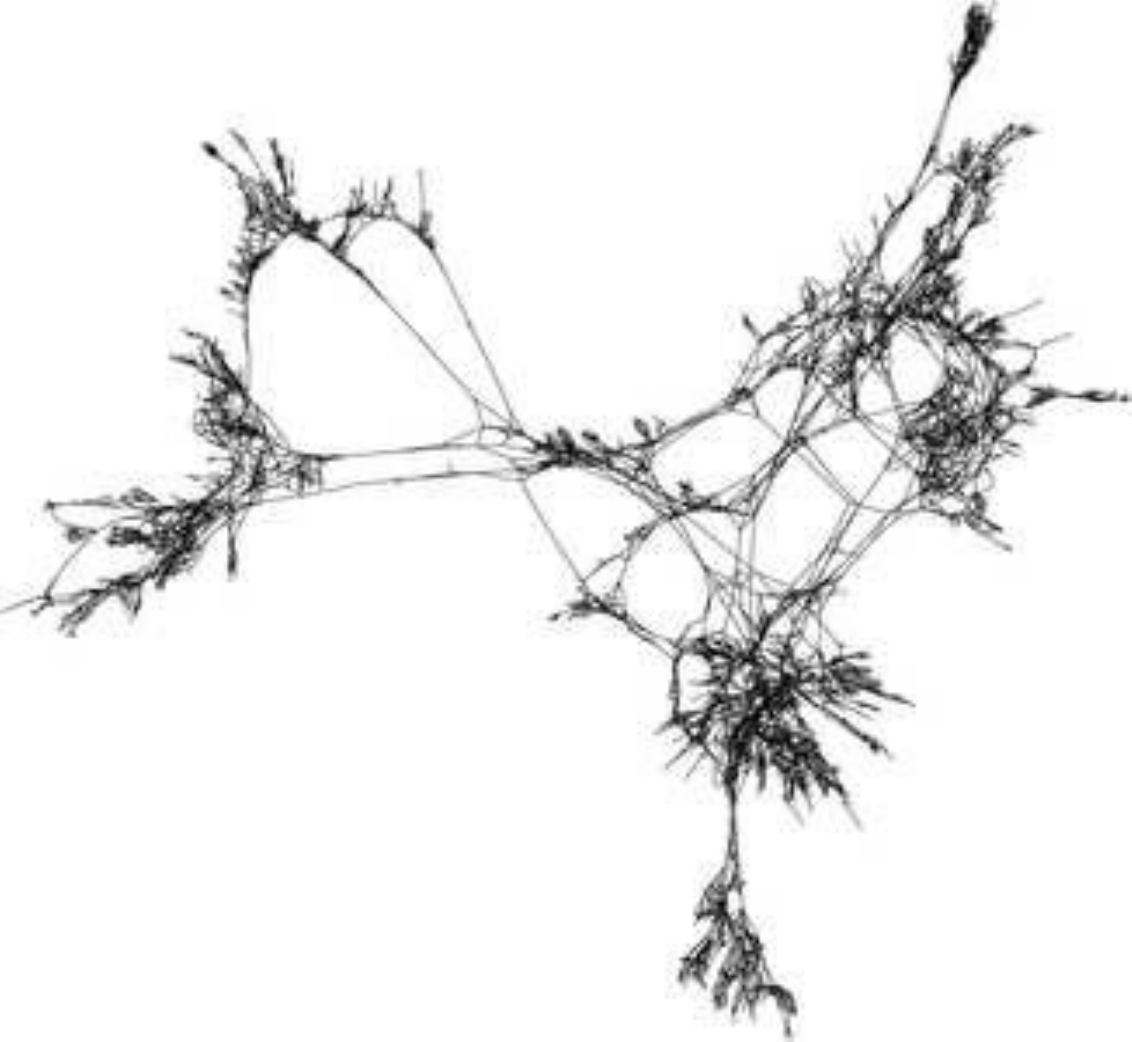}}
\caption{Examples of synthetic networks generated from electrical power grid of the western United States \cite{Watts1998}.
The synthetic networks seem to retain the global and local structure of the original power grid.}\label{fig:wattsexample}
\end{figure}

\begin{figure}[H]
\centering
\subfloat[Original network]{\label{fig:orig2}\includegraphics[width=0.35\textwidth]{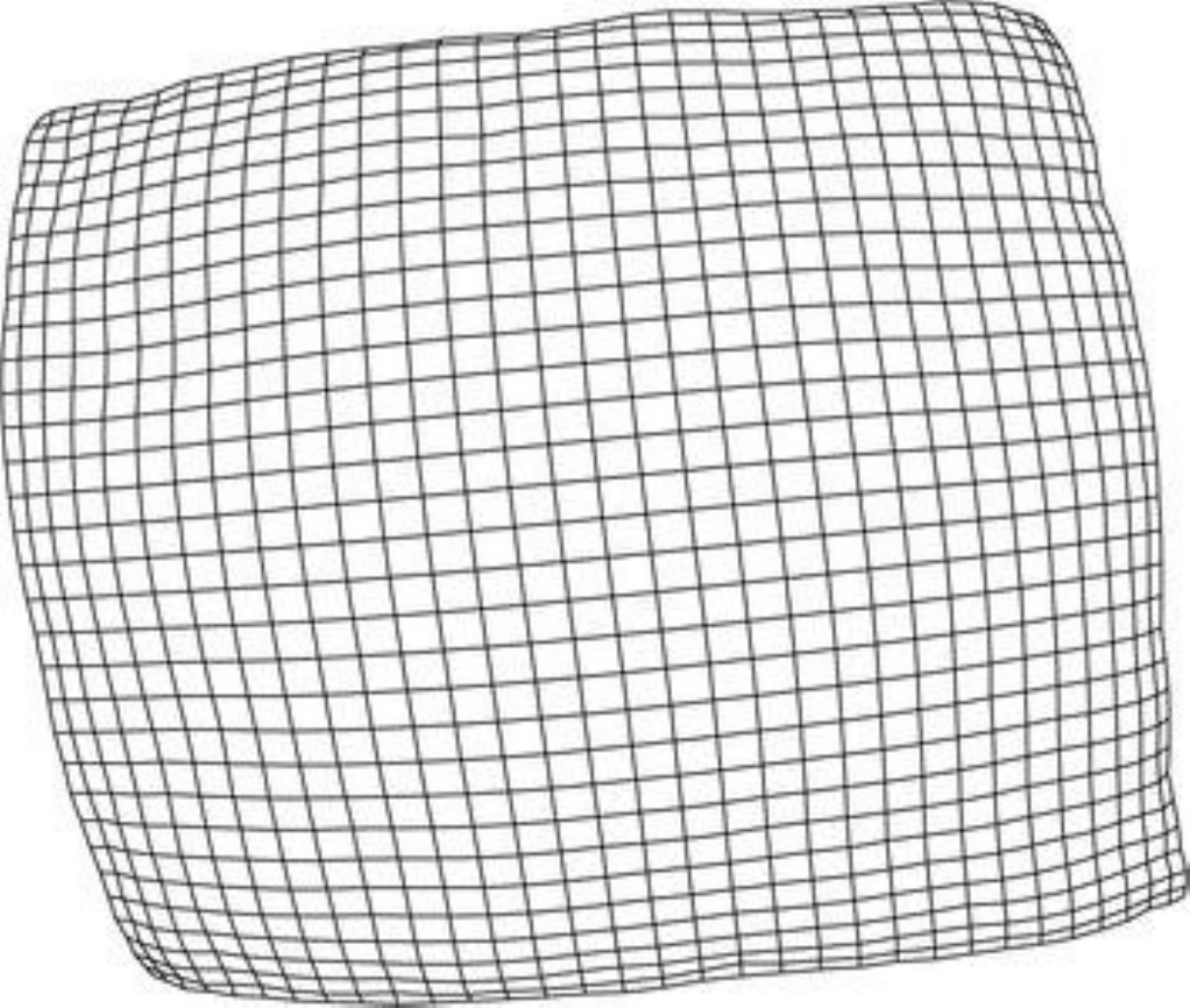}}
\subfloat[Fine level edge edits]{\label{fig:fine2}\includegraphics[width=0.35\textwidth]{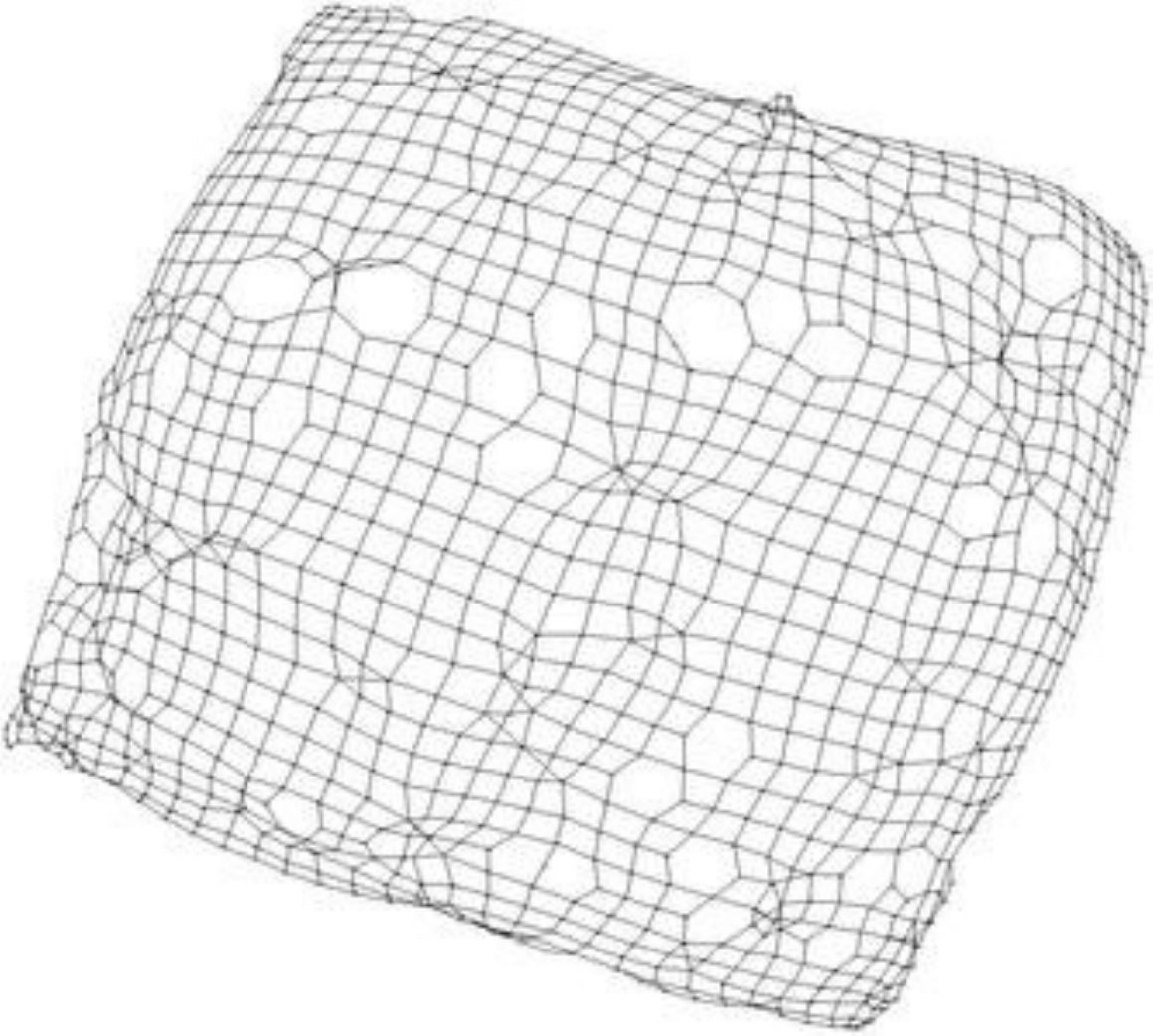}}\\
\subfloat[Small number of coarse edge edits]{\label{fig:coarse2}\includegraphics[width=0.35\textwidth]{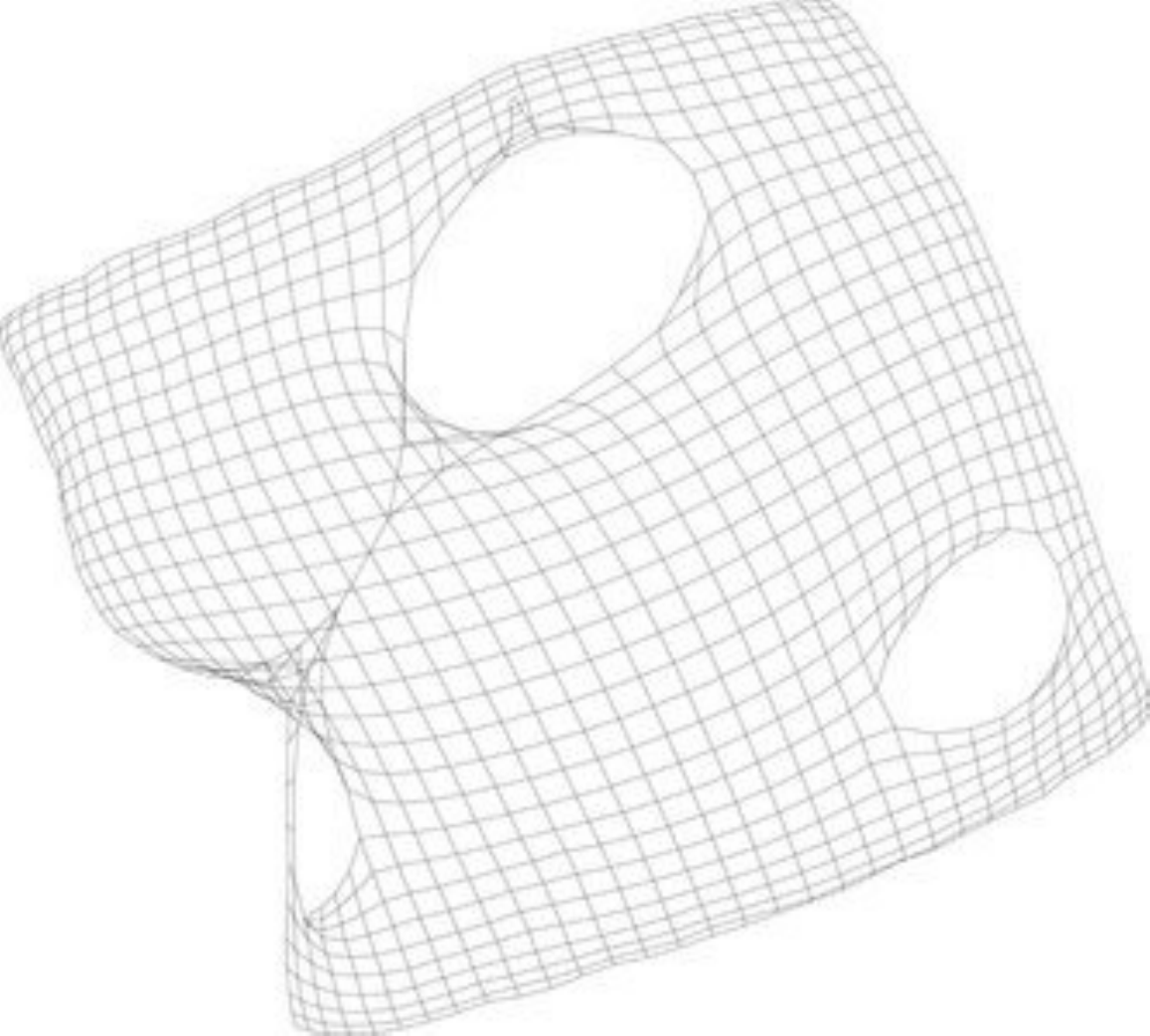}}
\subfloat[Fine edge edits with node growth]{\label{fig:finegrowth2}\includegraphics[width=0.35\textwidth]{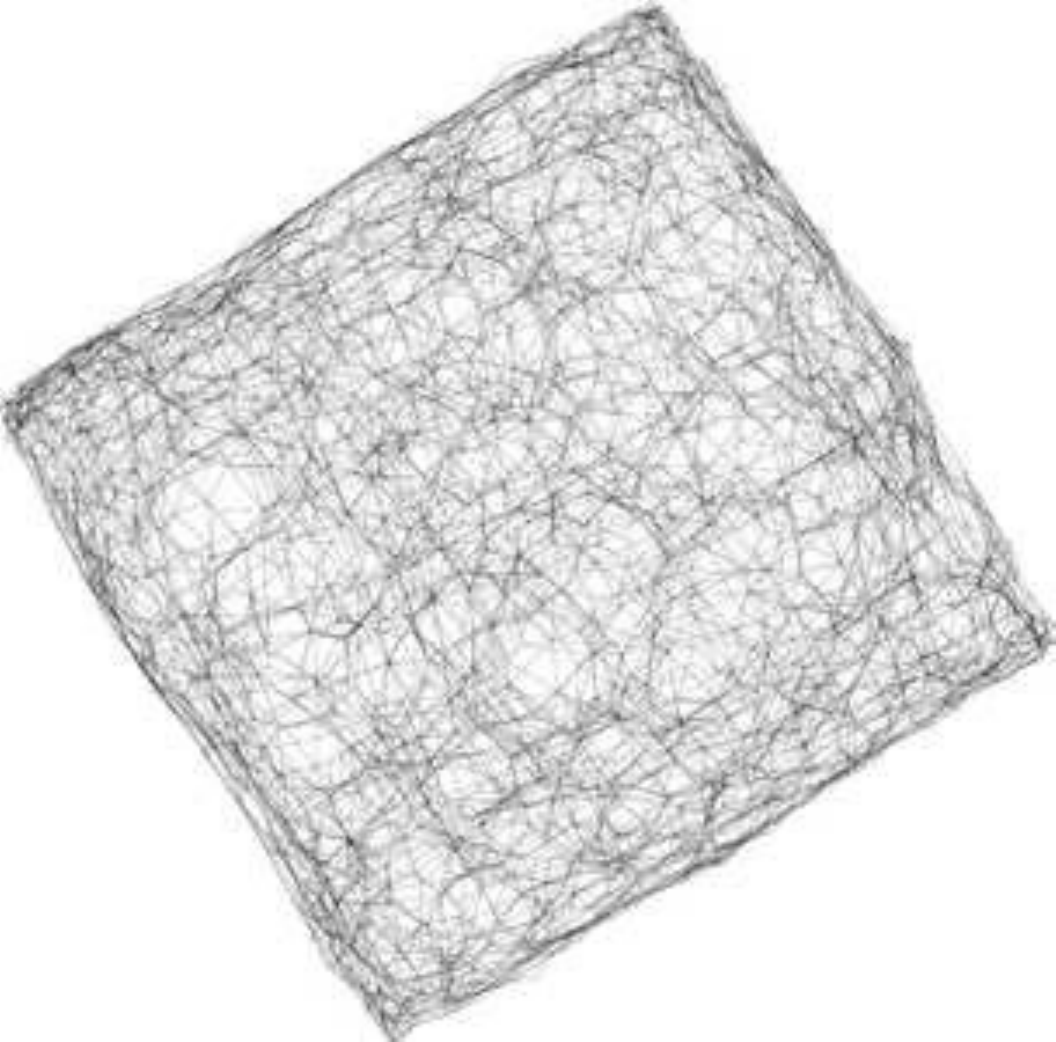}}\\
\subfloat[Several coarse edge changes]{\label{fig:globchanges2}\includegraphics[width=0.35\textwidth]{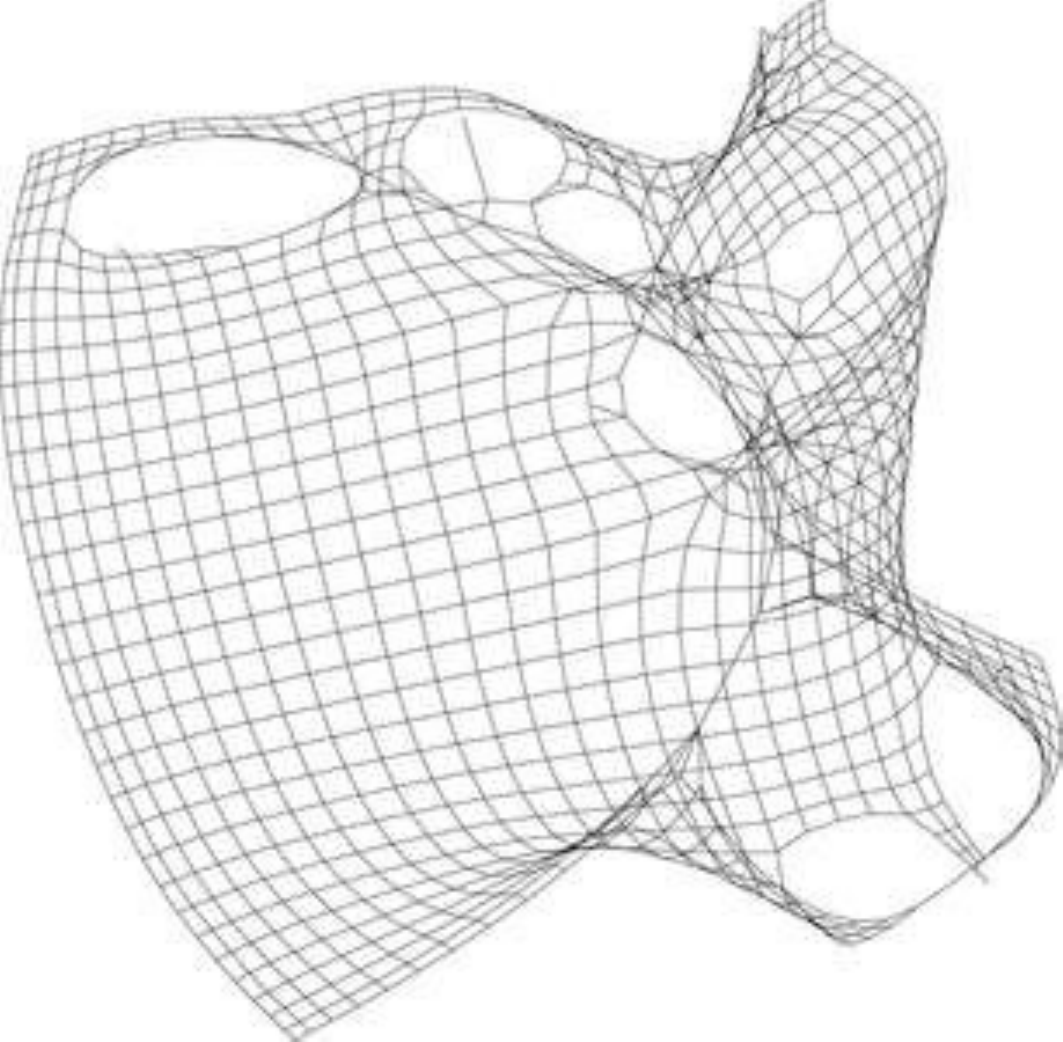}}
\subfloat[Coarse edge edits and node growth]{\label{fig:globgrowth2}\includegraphics[width=0.35\textwidth]{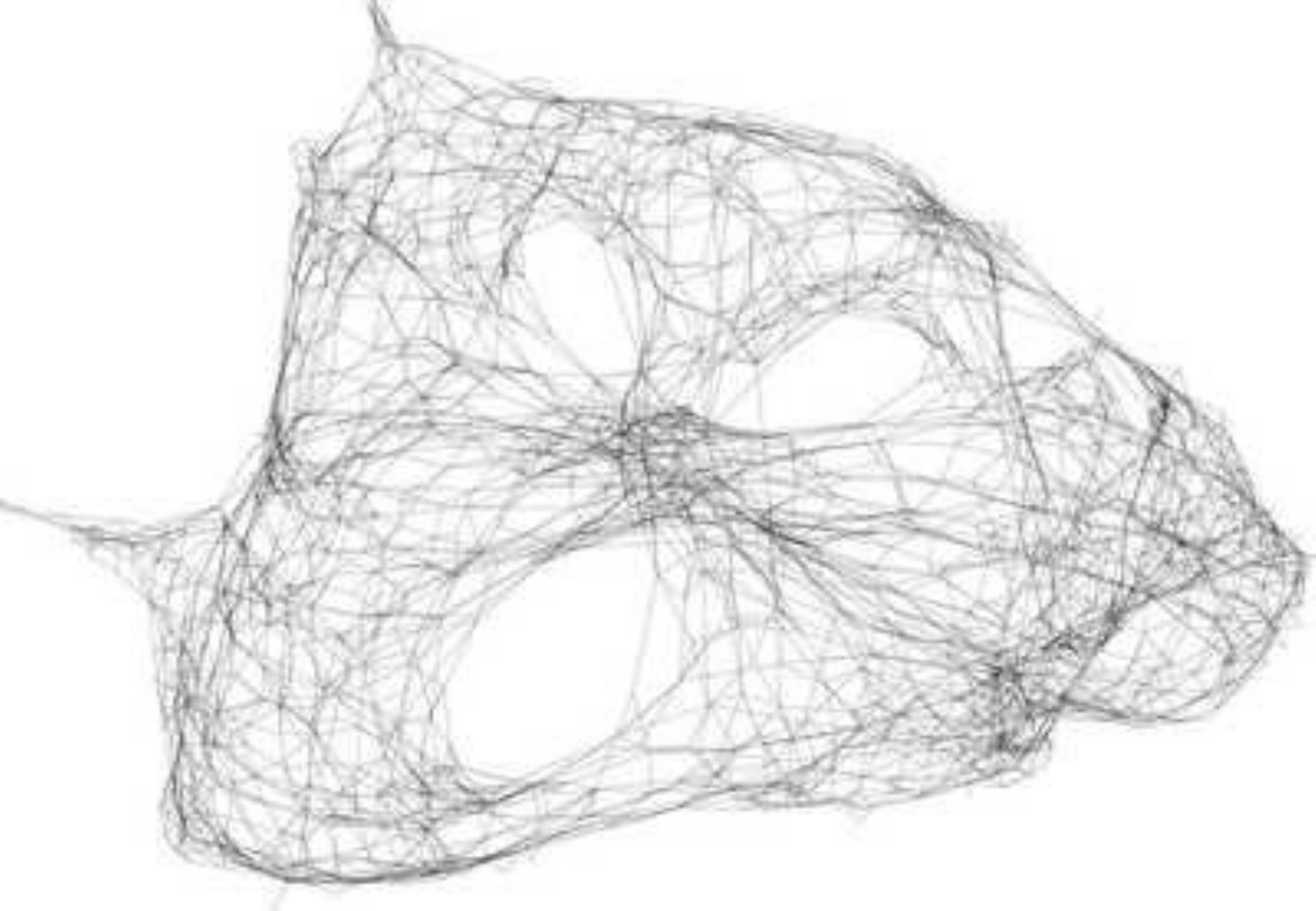}}
\caption{Examples of synthetic networks generated from mesh graph 33x33.}\label{fig:mesh33x33}
\end{figure}

\newpage

\bibliographystyle{plain}
\bibliography{netgen}

\vspace*{1cm}
\hspace*{1.5in}{\scriptsize\framebox{\parbox{2.4in}{
The submitted manuscript has been created in part by UChicago Argonne, LLC, Operator of Argonne National Laboratory (``Argonne'').  Argonne, a U.S. Department of Energy Office of Science laboratory, is operated under Contract No. DE-AC02-06CH11357.  The U.S. Government retains for itself, and others acting on its behalf, a paid-up nonexclusive, irrevocable worldwide license in said article to reproduce, prepare derivative works, distribute copies to the public, and perform publicly and display publicly, by or on behalf of the Government.
}}}

\end{document}